\documentclass[aip,jcp,amsmath,amssymb,twocolumn,reprint]{revtex4}

\usepackage{amsmath}
\usepackage{amssymb}
\usepackage{amscd}
\usepackage[pdftex]{graphicx}
\usepackage{nicefrac}
\usepackage[np,autolanguage]{numprint}
\usepackage{longtable}
\usepackage{bm}
\usepackage{float}
\usepackage[caption = false]{subfig}

\newcommand{\be}{\begin{equation}}
\newcommand{\ee}{\end{equation}}  
\newcommand{\bea}{\begin{eqnarray}}
\newcommand{\eea}{\end{eqnarray}}

\newcommand{\brho}{\boldsymbol{\rho}}
\newcommand{\br}{\textbf{r}}

\newcommand{\bz}{\textbf{z}}

\newcommand{\Eq}[1]{Eq.~\eqref{#1}}

\newcommand{\Fig}{Fig. \ref}

\begin{document}

\npthousandsep{}

\title{Assessing the Performance of the Diffusion Monte Carlo Method as Applied to the Water Monomer, Dimer, and Hexamer.}

\author{Joel D. Mallory}
\email{jdmallor@uci.edu}
\author{Sandra E. Brown}
\email{sebrown1@uci.edu}
\author{Vladimir A. Mandelshtam}
\email{mandelsh@uci.edu}
\affiliation{Department of Chemistry, University of California, Irvine, 1102 Natural
Sciences II Irvine, California 92697, USA} 

\date{\today} 

\begin{abstract}
The Diffusion Monte Carlo (DMC) method is applied to the water monomer, dimer, and hexamer, using q-TIP4P/F, one of the most simple, empirical water models with flexible monomers.
The bias in the time step ($\Delta\tau$) and population size ($N_w$) is investigated. For the binding energies, the bias in $\Delta\tau$ cancels nearly completely, while a noticeable 
bias in $N_w$ still remains. However, for the isotope shift, (e.g, in the dimer binding energies between (H$_2$O)$_2$ and (D$_2$O)$_2$) the systematic errors in $N_w$ do cancel. 
Consequently, very accurate results for the latter (within $\sim 0.01$ kcal/mol) are obtained with relatively moderate numerical effort ($N_w\sim 10^3$). For the water hexamer and 
its (D$_2$O)$_6$ isotopomer the DMC results as a function of $N_w$ are examined for the cage and prism isomers. For a given isomer, the issue of the walker population leaking out of 
the corresponding basin of attraction is addressed by using appropriate geometric constraints. The population size bias for the hexamer is more severe, and in order to maintain 
accuracy similar to that of the dimer, the population size $N_w$ must be increased by about two orders of magnitude. Fortunately, when the energy difference between cage and prism 
is taken, the biases cancel, thereby reducing the systematic errors to within $\sim 0.01$ kcal/mol when using a population of $N_w=4.8\times 10^5$ walkers. Consequently, a very 
accurate result for the isotope shift is also obtained. Notably, both the quantum and the isotope effects for the prism-cage energy difference are small.
\end{abstract}


\maketitle

\section*{Introduction}
\label{sect:intro}
Diffusion Monte Carlo (DMC)\cite{anderson1975,anderson1976,viel2001,mccoy2006} has engendered significant attention in the literature because it is one of the few numerical 
methods that enable the computation of the ground state of many-body systems. Large and cumbersome basis sets that scale exponentially with the system size and constitute an 
essential component of most variational methods are nonexistent in DMC\cite{austin2011}. Consequently, the computational cost of DMC increases rather slowly with the particle 
number\cite{mccoy2013_2}, which explains the method's attractiveness for treatments of systems with many degrees of freedom.

DMC is appealing for theoretical studies of condensed matter systems and has developed a reputation for its widespread application to weakly-bound ensembles of molecules that 
interact through hydrogen bonding and dispersion forces\cite{jones2009,babin2013}. Some examples of systems that have been examined with DMC are Bose condensates of parahydrogen 
and helium\cite{boninsegni2012,cuervo2005,guardiola2008,sola2011}; sheets of graphite and diamond\cite{nemec2010}; fcc crystallized xenon at 0 K\cite{jones2009}; HF, HCN, and SF$_6$ 
trapped in clusters of argon and helium\cite{viel2001,bacic2005}; and water clusters with a special emphasis placed on the water 
hexamer\cite{wang2012,buch1999,buch2003,goldman2004,gillan2013,gregory1996_1,gregory1996_2,clary1996}.

As originally proposed by Anderson\cite{anderson1975,anderson1976} DMC takes advantage of the similarity between the imaginary time-dependent Schr\"{o}dinger equation and the 
diffusion equation. The method employs a population of random walkers (or $\delta$-functions) that evolves in imaginary time $\tau$ and samples the configuration space to  
collectively represent the lowest energy eigenstate of the system defined by a potential energy surface (PES).  

A longstanding issue with DMC is that it suffers from inherent sources of systematic errors that arise due to the use of finite values for the time step $\Delta\tau$ and population 
size $N_w$, as well as the need to introduce a population control mechanism in the form of branching steps\cite{boninsegni2012,cuervo2005,assaraf2000,jones2009}. We have found that 
an analysis of the behavior and extent of the biases is often absent or ignored altogether when the results from DMC simulations are reported (see, e.g., 
refs.~\onlinecite{wang2012,buch1999,buch2003,gregory1996_1,guardiola2008,sola2011}). Ideally, the information on the DMC energy estimate at finite values of $\Delta\tau$ and $N_w $ can be 
analyzed and extrapolated to $\Delta\tau\to 0$ and  $N_w \to\infty$. Yet, ref.~\onlinecite{boninsegni2012} suggests that the bias of the DMC energy estimate, specifically, the 
population size bias, can be very nontrivial and as such is often difficult to extrapolate.

Nevertheless, efforts have been underway for some time to substantially reduce or even eliminate the biases with respect to the time step and population size. The price to be paid, 
however, is that the algorithms are more complicated than the simple version developed by Anderson. These variants of DMC usually implement importance sampling (IS-DMC) whereby a 
drift velocity term is incorporated into the imaginary time Schr\"{o}dinger equation to drive the random walkers into regions of configuration space where the wavefunction is largest. 
This procedure often utilizes an optimized guiding or trial wavefunction\cite{watts1991,umrigar1993,assaraf2000,warren2006,sola2011}.

The time step error was studied extensively by Umrigar in ref.~\onlinecite{umrigar1993}. It was determined that the bias could be considerably diminished if diffusive 
displacements of the random walkers are accepted or rejected based on a ratio of values for the trial wavefunction at the current and previous points in configuration space. 
This methodology can only be implemented in conjunction with IS-DMC, as the probability for acceptance of the diffusion move is directly linked to a Gaussian distribution containing the 
aforementioned drift term. It was demonstrated that this algorithm does, in principle, reduce the time step error, but only if an accurate form for the trial wavefunction is known 
\textit{a priori}\cite{umrigar1993}. Thus, the most practical approach to remove the time step error is still to run several simulations at different values of $\Delta\tau$ and extrapolate 
the result to $\Delta\tau=0$.

An ``improved DMC method'' proposed in ref.~\onlinecite{assaraf2000} establishes a fixed population of random walkers with unequal weights. Spurious correlations among the otherwise 
independent motions of the walkers contribute to the population size (or population control) bias and arise when the weights are reconfigured during the branching steps. Therefore, 
stochastic reconfigurations are carried out with the intention of minimizing the fluctuations in the weights as much as possible. This technique may decrease the population size bias 
and the statistical fluctuations in the energy\cite{assaraf2000,nemec2010}.

Another DMC variant, called norm-conserving DMC, exactly balances the flux of walkers entering and exiting the ensemble such that the population remains constant for all iterations of 
the simulation\cite{jones2009}. Weights for the walkers are not introduced except through a mean-field scheme that is invoked to decrease the population size bias, which seems to scale 
as $\sim 1/N_w $. It was asserted that this approach could entirely eliminate the population size bias, but in reality the bias will vanish strictly in the limit of 
$N_w \rightarrow\infty$\cite{jones2009,assaraf2000,nemec2010,hetherington1984,boninsegni2012}.

The issue of a non-vanishing population size bias was reported by Boninsegni\cite{boninsegni2012} upon observing discrepancies between DMC energies for helium clusters and energies obtained 
through an alternative method\cite{cuervo2005}. In a subsequent study of parahydrogen clusters, the bias was thoroughly investigated and found to never completely disappear, even in an attempt 
to extrapolate the result to $N_w \rightarrow\infty$. Indeed, the dependence of the energy on the reciprocal walker number proved to be nontrivial, and the resultant curve seemed to diverge in 
the limit of large walker numbers. In light of this study we believe that the population size bias must be thoroughly examined whenever DMC is utilized in treatments of many-body systems. The 
suggested divergence of the DMC energy at large $N_w$ should make all reported DMC results questionable, whenever such a study is lacking. 

Furthermore, ref.~\onlinecite{boninsegni2012} utilizes the DMC variant outlined in ref.~\onlinecite{umrigar1993} which requires IS-DMC such that the population size bias happens to be quite 
sensitive to the choice of the trial wavefunction. If a substantial discrepancy exists between the trial and exact ground state wavefunctions, the systematic bias with respect to $N_w$ becomes 
extremely pronounced. Since such a discrepancy is unavoidable, the pathological convergence with respect to the number of random walkers may be an intrinsic property of IS-DMC. In other words, 
IS-DMC may not be ideal for treating complex systems with complicated ground state wavefunctions, as any attempt to optimize the trial wavefunction by finding a suitable parametrizaion is bound 
to fail. Water clusters seem to provide a good example of such systems, in which the complexity is due to the existence of two different time scales corresponding to the slow inter- and fast 
intra-molecular degrees of freedom.

The water hexamer has been intensely studied from many different experimental and theoretical angles because it represents the smallest cluster of water molecules that still maintains a 
three-dimensional configuration\cite{clary1996,wang2012,babin2013,gregory1996_1,goldman2004}. This fundamental building block of water and ice retains many of the physical 
properties of bulk water and the ability to exist in discrete isomeric forms. In order to avoid dealing with the two time scales most of the DMC simulations of water clusters and, in 
particular, the water hexamer have invoked the frozen monomer approximation such that the intramolecular degrees of freedom for the constituent water molecules were collectively 
neglected\cite{buch1999,buch2003,gregory1996_1,gregory1996_2,goldman2004}. This approach assumes that the intra- and inter-molecular motions can be adiabatically decoupled, and as such, 
lead to a substantial simplification of the water dynamics and its numerical treatment. Unfortunately, this approximation fails to describe many important properties of water, including 
the notable isotope effect due to the substitution of hydrogen atoms by deuterium atoms. 

Consequently, in the present study we consider a flexible water model in conjunction with full-dimensional DMC. Nonetheless, our focus is not to carry out accurate first-principles simulations 
of water that would, in particular, employ an ``accurate'', possibly {\it ab initio}, water PES. Instead, we focus our attention on the methodological issues in an attempt to assess the performance 
and various other aspects of DMC implementation for a complex many-body system---a prime example of which is the water hexamer. In order to make our goal feasible, we minimize the computational cost 
by employing the q-TIP4P/F PES; one of the least expensive empirical, flexible water models available. This potential has proved to be reasonable in describing thermodynamic properties of water 
with quantum nuclear effects included\cite{habershon2009}. We note that presumably more accurate PES's exist, such as the WHBB\cite{wang2011} or HBB2-pol\cite{medders2013}, and the most recent 
and most accurate MB-pol PES\cite{babin2014}, which are all {\it ab initio}-based. Unfortunately, all these potentials are very expensive, and by orders of magnitude more expensive than q-TIP4P/F. 
We believe that the present study using the inexpensive PES will allow us to carry out DMC simulations employing sufficiently long projection times and large population sizes to be able to assess 
the methodology before it is applied to more realistic, albeit more expensive, PES's.

Notably, Bowman and co-workers\cite{wang2012} have already reported DMC results on the ground state energies of three isomers of the WHBB water hexamer (cage, prism, and book). The authors briefly 
acknowledge in the supplemental information that a strong dependence of the isomer energies on the population size is evident, but the extent and behavior of the bias was not rigorously characterized. 
When taking the energy difference, the large systematic errors may or may not cancel, and a nontrivial bias can arise for the energy difference with respect to the population size and/or time step. 
Most importantly, however, the issue arising due to leakage of the random walkers out of their original basin of attraction has not been thoroughly addressed. We have observed that the previous DMC 
treatments of the water hexamer using rigid monomers\cite{buch2003,gregory1996_1,gregory1996_2,goldman2004} either did not consider this problem, or did not consider it adequately, but we are unaware 
of whether it poses a serious problem for clusters comprised of rigid water monomers. For example, ref.~\onlinecite{gregory1996_1} does mention this problem, but makes an impression that it is not 
serious. The authors of the latter work suggested that using a simple constraint based on monitoring only the energies of random walkers solves the problem. In the case of a flexible water hexamer (at least 
with the q-TIP4P/F PES), the migration of the population out of the basin of attraction in question occurs for all the isomers considered in our test calculations (some of which are not reported here). Therefore, a numerically inexpensive and effective 
solution to this problem appears to be nontrivial. In particular, a simple energy-based constraint would fail, because the energy differences between different isomers (and the corresponding quantum energies) are small. In contrast, the 
authors of ref.~\onlinecite{bacic2005} proposed the use of a relatively simple geometric constraint that seemed to prevent the population migration in their DMC simulation of Ar$_n$HF van der Waals 
clusters. Our approach involves implementation of more than one simple geometric constraint.
 
In what follows, we present a comprehensive assessment of the performance of DMC as prescribed 
by Anderson\cite{anderson1975,anderson1976} and its sources of systematic bias in applications to 
the water monomer, dimer, and hexamer. In addition, we consider the isotope substitution of hydrogens by deuteriums and, consequently, investigate the isotope shifts and their dependence on both 
the population size and the time step. 

\section*{The Diffusion Monte Carlo Method}
\label{sect:comput}
Consider an $N$-particle system described by the Hamiltonian 
\begin{equation}
\hat H = \hat T + V(\br ) = \sum_{i=1}^{3N}\frac{\hat{p_i}^2}{2m_i}+ V(\br ) 
\label{Hsys}
\end{equation}
where $V(\br )$ defines the PES, $m_i$ are particle masses, and 
$\br=(\br_1,...,\br_{3N})\in\mathbb{R}^{3N}$ is the coordinate vector.

The DMC method of Anderson exploits the apparent isomorphism between the imaginary time-dependent Schr\"{o}dinger equation 
and the standard diffusion equation\cite{anderson1975,anderson1976} 
\be\label{eq:diff}
\frac{\partial\Psi}{\partial\tau}=-(\hat{H}-E_{\rm ref})\Psi=-\hat T \Psi-(\hat{V}-E_{\rm ref})\Psi
\ee
with the constant shift $E_{\rm ref}$ defining the energy reference. This equation can be solved and recast in terms 
of the imaginary time propagator 
\be\label{eq:prop}
\Psi(\br;\tau)=e^{(E_{\rm ref}-\hat{H})\tau}\Psi(\br,0)
\ee
By expanding $\Psi(0)$ in the eigenbasis of $\hat{H}$ 
and substituting into \Eq{eq:prop},
\bea\nonumber
\Psi(\br;\tau)&=&\sum_{k}c_k e^{(E_{\rm ref}-E_k)\tau}\psi_k(\br)\\\label{eq:expand}
&=&e^{(E_{\rm ref}-E_0) \tau}\sum_{k}c_k e^{(E_0-E_k)\tau}\psi_k(\br) 
\eea
one can see that
in the $\tau\to\infty$ limit $\Psi(\br;\tau)$ becomes proportional to the ground state wavefunction $\Psi_0$,
as all the other contributions tend to zero exponentially, relative to the contribution of the ground state.
Furthermore, \Eq{eq:expand} also implies that
\be
\lim_{\tau\to\infty} \|\Psi\|=\left\{\begin{array}{lr} 0,&  \mbox{if}\ E_{\rm ref} <E_0\\
\infty ,&  \mbox{if}\ E_{\rm ref} >E_0\\
\mbox{const},&  \mbox{if}\ E_{\rm ref} =E_0 \end{array} \right.
\ee
That is, the solution of \Eq{eq:diff} is stationary only when $E_{\rm ref}$ coincides with the
ground state energy $E_0$.

\Eq{eq:diff} can be solved numerically by introducing a finite time step $\Delta\tau$ 
and invoking the split-operator approximation for the short-time propagator
\be
\Psi(\br;\tau+\Delta\tau) \approx 
e^{(E_{\rm ref}-\hat{V})\Delta\tau} e^{-\hat{T}\Delta\tau}\Psi(\br;\tau)
\ee
Assuming the wavefunction $\Psi$ is non-negative at all positions, it can be represented by an ensemble of 
delta functions $w_j\delta(\br-\br^{(j)})$ or random walkers $\br^{(j)}=\br^{(j)}(\tau)$  ($j=1,...,N_w$) that 
evolve in time $\tau$ together with their weights $w_j=w_j(\tau)$. 

Operating with the kinetic energy propagator on the $i$-th component of the $j$-th $\delta$-function,
\[
\delta(\br-\br^{(j)})=\prod_{i=1}^{3N}\delta(\br_i-\br_i^{(j)})
\]
gives a Gaussian distribution that governs the diffusive displacements for a single component of the $j$-th random walker 
in configuration space 
\begin{equation}
\exp \left[-\frac{\Delta\tau\hat{p_i}^2}{2m_i}\right]\delta(\br_i-\br_i^{(j)})=
\sqrt{\frac{m_i}{2\pi\Delta\tau}}\exp \left[-\frac{m_i(\br_i-\br_i^{(j)})^2}{2\Delta\tau}\right]
\end{equation}
Consequently, the kinetic energy part of the random walker propagation can be represented by shifting the $i$-th component of $j$-th walker randomly,
\be\label{eq:diff1}
\br_i^{(j)}(\tau+\Delta\tau)=\br_i^{(j)}(\tau)+\bz_i
\ee
where $\bz_i$ is a Gaussian random number with standard deviation
\[
\sigma_i=\sqrt{\frac{\Delta\tau}{m_i}}
\]

The action of the potential energy propagator on the walker population can be implemented in many different ways. The most straight-forward recipe is to multiply 
each weight $w_j$ by the factor 
\be
p_j=\exp\Big[\Big(E_{\rm ref} -V\big(\br^{(j)}\big)\Big)\Delta\tau\Big] 
\ee
However, this will soon make the random walker representation of the wavefunction inefficient due to the appearance of random walkers with either very small or very 
large weights. Here we adapt the simplest version of DMC, in which the weights are all the same, $w_j\equiv 1$, but upon each advancement by $\Delta\tau$ a ``branching'' 
procedure is implemented, in which some walkers are replicated and some are killed. 

The value of of $p_j$ computed upon each diffusion displacement determines the number of copies ($n_j$) of the random walker (i.e. $\br^{(j)}$) 
that will be retained in the ensemble. Precisely,
\be
n_j=\left\{\begin{array}{rl} \left[p_j\right]+1,&  \mbox{if}\ \left\{p_j\right\}>\xi \\
\left[p_j\right]  ,&  \mbox{otherwise}\end{array} \right.
\ee
where $\xi\in [0,1]$ is a uniformly distributed random number. In particular, the random walker is killed when $n_j=0$.

Upon completion of each branching step the value $E_{\rm ref}$ is updated to reflect changes in the population size $N_w$ according to 
\begin{equation}
E_{\rm ref}(\tau)=\bar{V}(\tau)-\alpha\frac{N_w (\tau)-N_w (0)}{N_w (0)}
\end{equation}
where 
\be 
\bar{V}(\tau)= \frac 1 {N_w(\tau)} \sum_j V( \br^{(j)}(\tau))
\ee
is the average of the potential energy over the random walker population, $N_w (\tau)$ is the instantaneous number of random walkers present in the population, and $\alpha$ is 
a proportionality constant, usually chosen to be $1/\Delta\tau$. This (stabilization) procedure is imposed to keep the population size stationary in time throughout 
the course of simulation. Otherwise, without the stabilization, the population would explode or crash exponentially. Likewise, the value of $E_{\rm ref}$ fluctuates in time $\tau$ 
in a stationary manner, and its time average can be used to estimate the ground state eigenenergy $E_0$. Depending on the initial conditions for the random walker population, the 
stationarity is achieved only at sufficiently long projection times. 

One should not be mislead by \Eq{eq:expand}, which gives the impression that the convergence of the present version of the DMC algorithm is defined by the maximum projection time $\tau_{\rm max}$ according to the rates with which all the unwanted contributions vanish exponentially (relative to the contribution of the ground state $\Psi_0$) due to the factors $e^{(E_0-E_k)\tau_{\rm max}}$. This would be the case if the ground state energy was estimated using the wavefunction $\Psi(\br;\tau)$ explicitly. However, in the present algorithm it is estimated by averaging $E_{\rm ref}$, which fluctuates, and the convergence of the DMC energy estimate is rather defined by the statistical errors associated with this average.

Aside from the modifications associated with geometric constraints specific for the system in question, the present DMC algorithm exactly resembles the algorithm developed by 
Anderson\cite{anderson1975,anderson1976}, which was also implemented later by a number of groups (see, e.g. refs.~\onlinecite{mccoy2013_2,mccoy2013_1,mccoy2006,watts1991,bowman2008,wang2012}). 
Following ref.~\onlinecite{bacic2005} we have augmented the branching step in order to prevent random walkers from escaping out of the basin of attraction associated with a chosen isomer. 
During the course of the simulation any walker that violates one of the imposed geometric constraints is eliminated from the ensemble. Ideally, such constraints must be inexpensive 
to evaluate compared to the cost of the potential energy calculation, yet they should be effective in separating the isomer in question from the rest of the configuration space. 
The geometric constraints that we have imposed for simulations involving the water hexamer (see below) require calculations of the pairwise distances between the oxygen atoms, distances 
of the oxygens from the center of mass, and the moments of inertia. Threshold values for the constraints are chosen as a compromise between restricting the configuration space only to 
the relevant basin of attraction and ensuring that all regions of the configuration space where the corresponding wavefunction has an appreciable amplitude are accessible by the walkers. 

However, one must always keep in mind that there is only one true ground state, while the physical meaning of any other ``ground state'' of the system, subject to certain geometric constraints, 
must be established independently of the simulation algorithm. This problem is intimately related to the problem of defining an ``isomer'', namely, a species that has a 
distinct structure, naturally associated with its own, local basin of attraction, a region in space that is unique enough to be separated from the other regions of configuration space by sufficiently 
large potential energy barriers. Alternatively, we can define an ``isomer'' as a state of the system, that once prepared, would exist for a sufficiently long time. Nevertheless, the actual stability of a quantum mechanical ground state restricted to a basin of attraction may not be trivial to establish, 
even approximately, without performing quantum dynamics simulations that are usually much more demanding than any ground state computation.

\section*{Water Monomer and Water Dimer}
\label{sect:monodimer}
 A series of calculations with varying time step $\Delta\tau$ and population size $N_w=N_w(0)$ have been performed for each of the following: H$_2$O, (H$_2$O)$_2$, D$_2$O, and (D$_2$O)$_2$. Namely, 
for a fixed value of $\Delta\tau=10$ au we ran six simulations with $N_w=2 \times 10^2,\ 4 \times 10^2,\ 9 \times 10^2,\ 1.6 \times 10^3,\ 3.6 \times 10^3,\ 1.96 \times 10^4$, and for fixed 
$N_w=1.96 \times 10^4$, five simulations with $\Delta\tau=2,\ 5,\ 10,\ 15,\ 20$ au. Given $\Delta\tau$ and $N_w$, 10 independent DMC simulations were usually carried out. For each run 
a relatively short equilibration (if necessary)  was performed before starting to accumulate the average energy $\langle E_{\rm ref}\rangle$. Consequently, the 10 energy averages were used to estimate the 
statistical error. The total projection time (i.e. including all 10 independent runs) for each $\Delta\tau$ and $N_w$ was $\tau_{\rm max}=2\times 10^6$ au.

\begin{figure}
\caption{Relative statistical errors for the ground state energy estimates for H$_2$O monomer and dimer as a function of $1/\sqrt{N_w }$ for a total projection time of $2.0 \times 10^6$ au. The 
errors were estimated by splitting the overall simulation into 10 independent simulations.}
\label{fig:errorbars}
\includegraphics[width=0.495\textwidth]{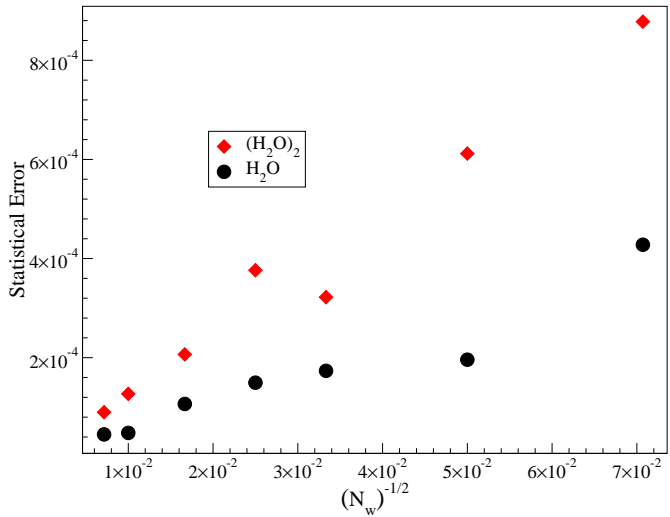}
\end{figure}

\begin{figure*}
\includegraphics[width=0.495\textwidth]{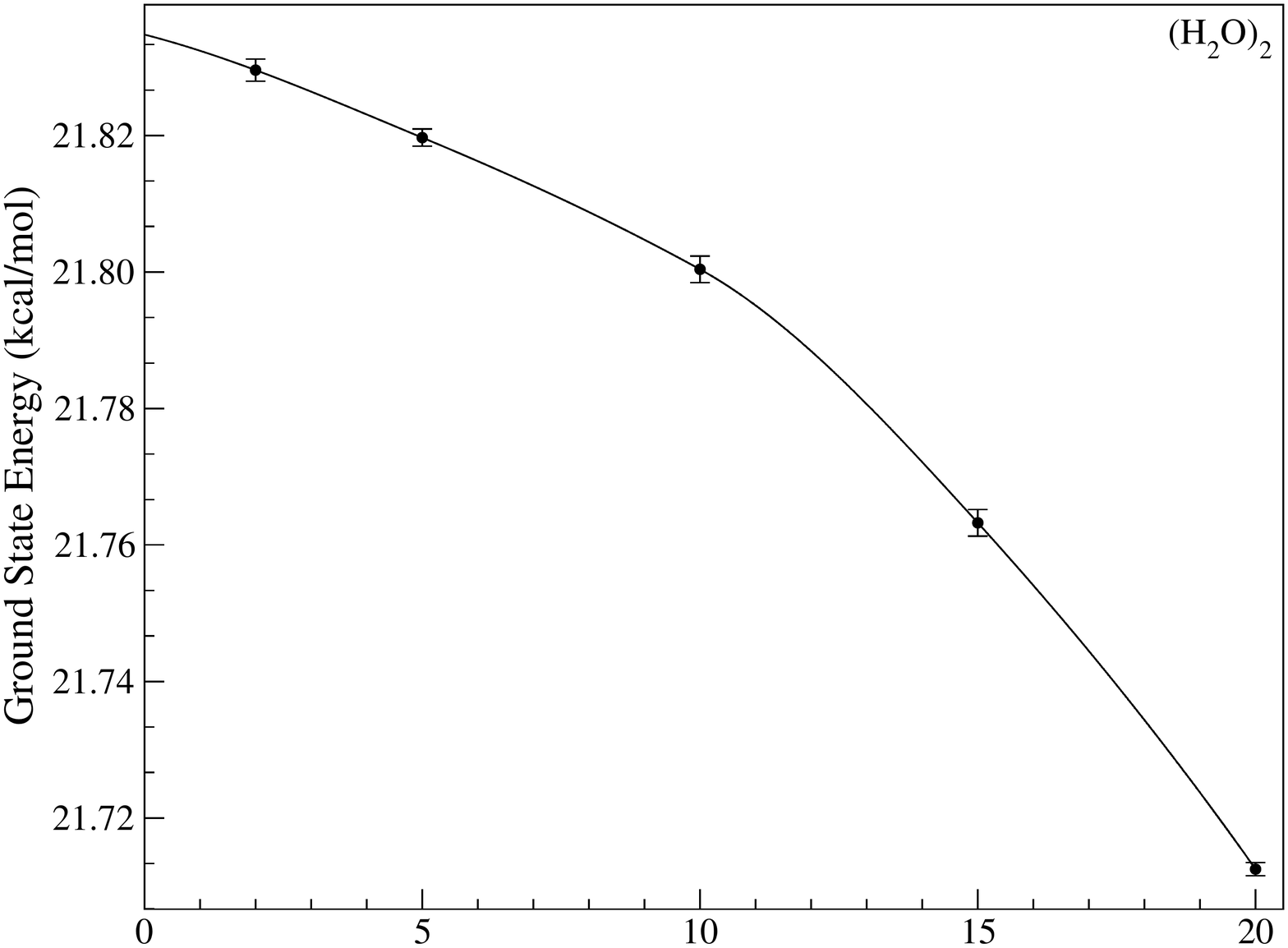}
\includegraphics[width=0.495\textwidth]{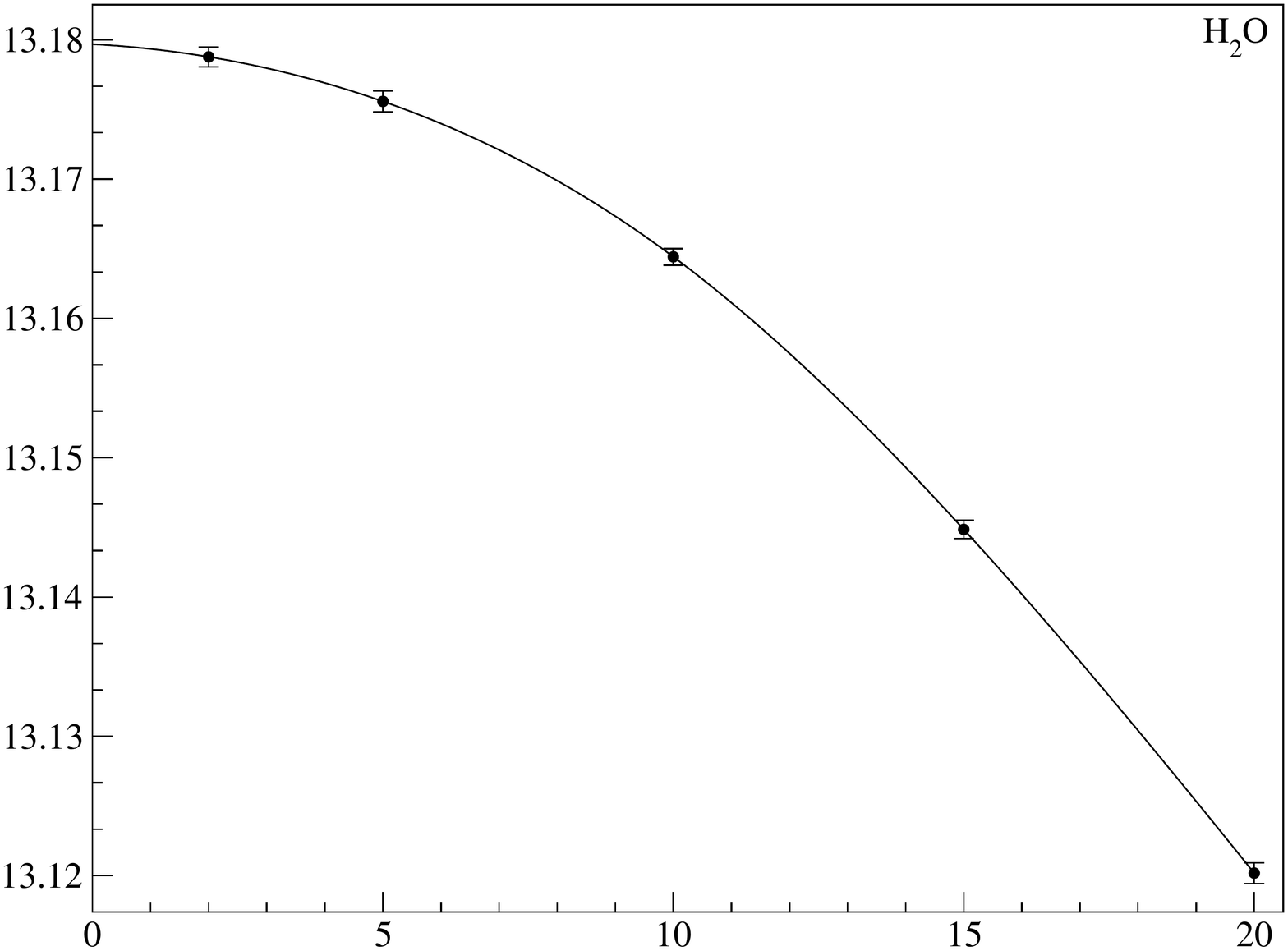}
\includegraphics[width=0.495\textwidth]{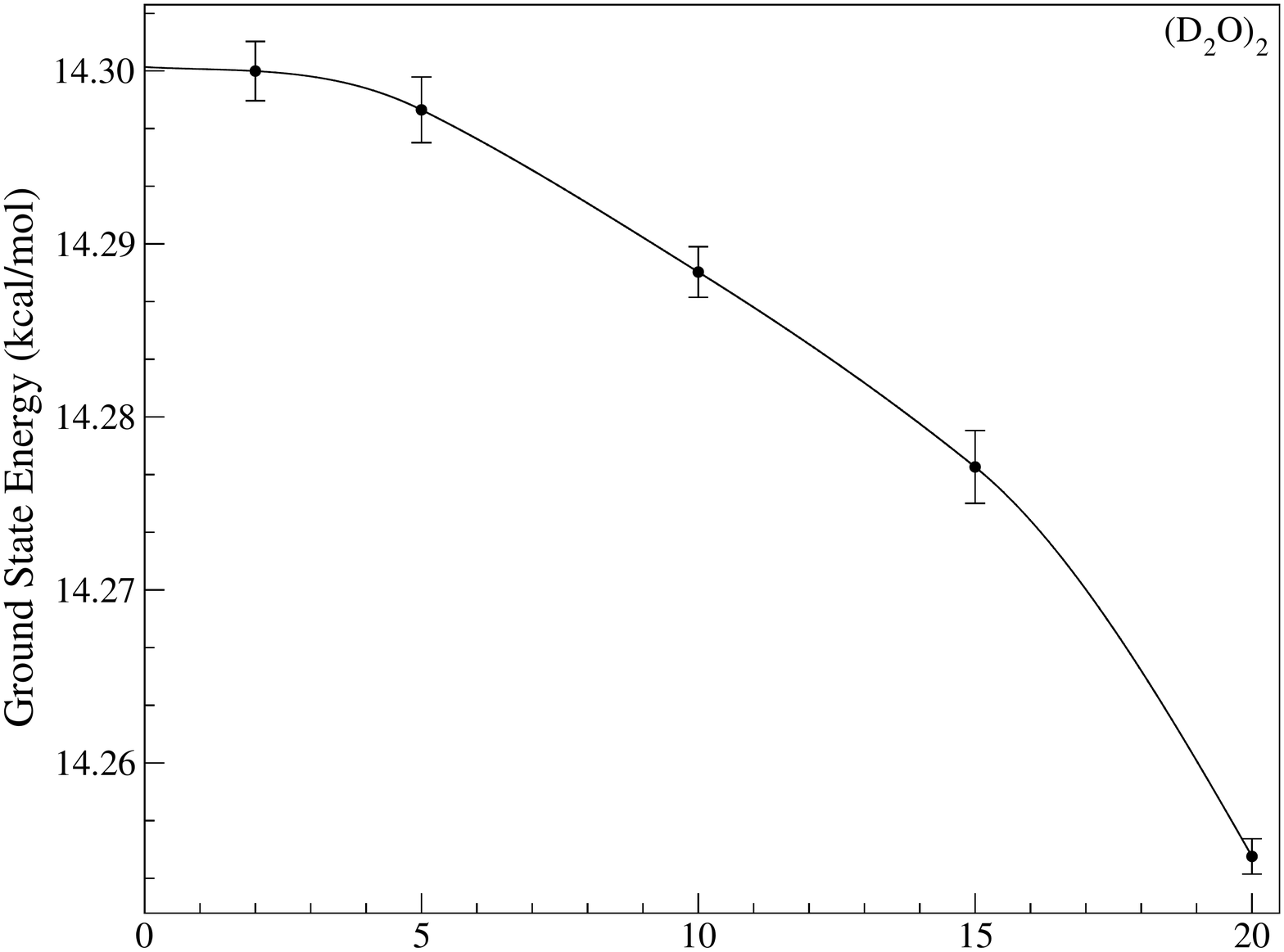}
\includegraphics[width=0.495\textwidth]{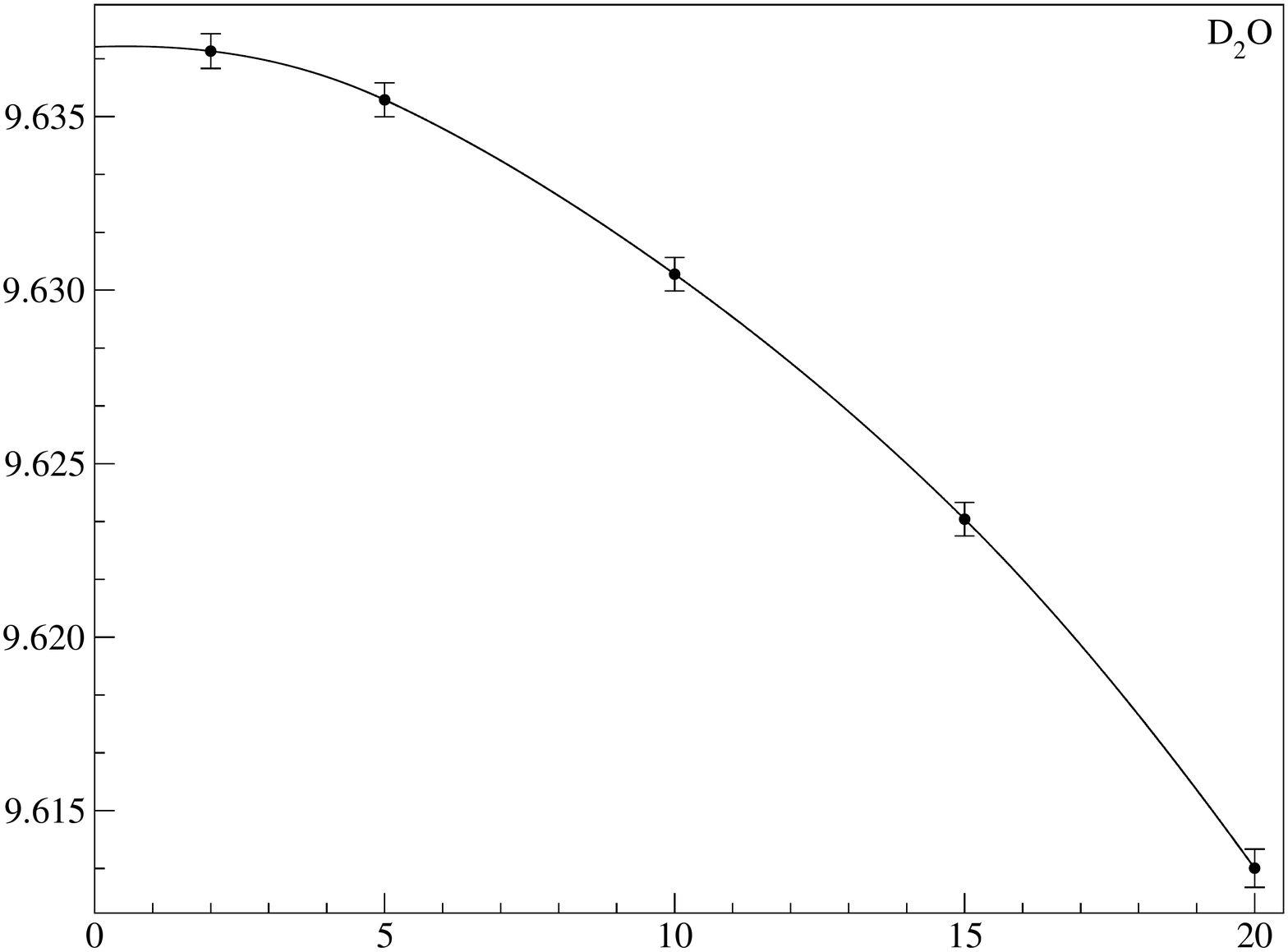}
\includegraphics[width=0.495\textwidth]{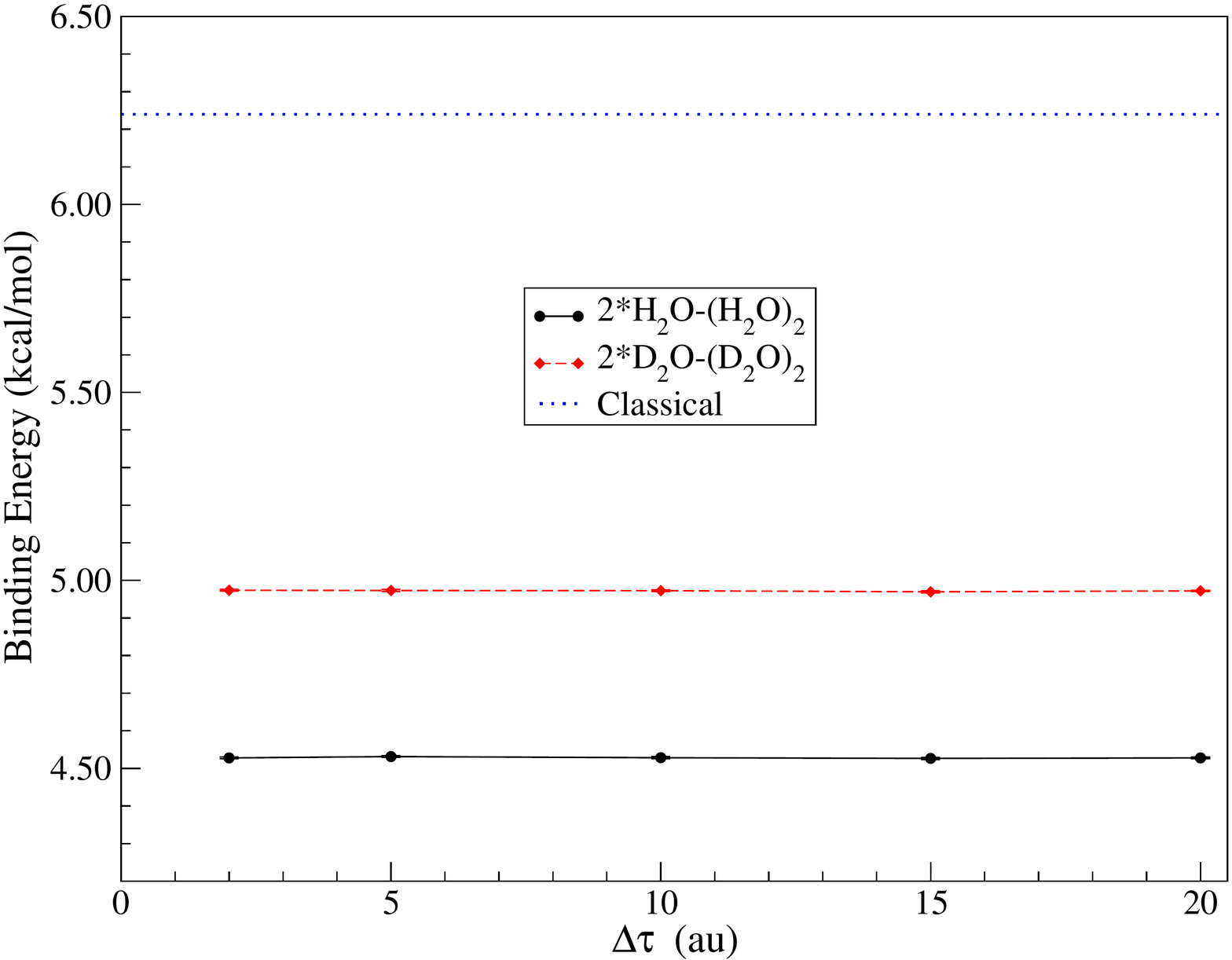}
\includegraphics[width=0.495\textwidth]{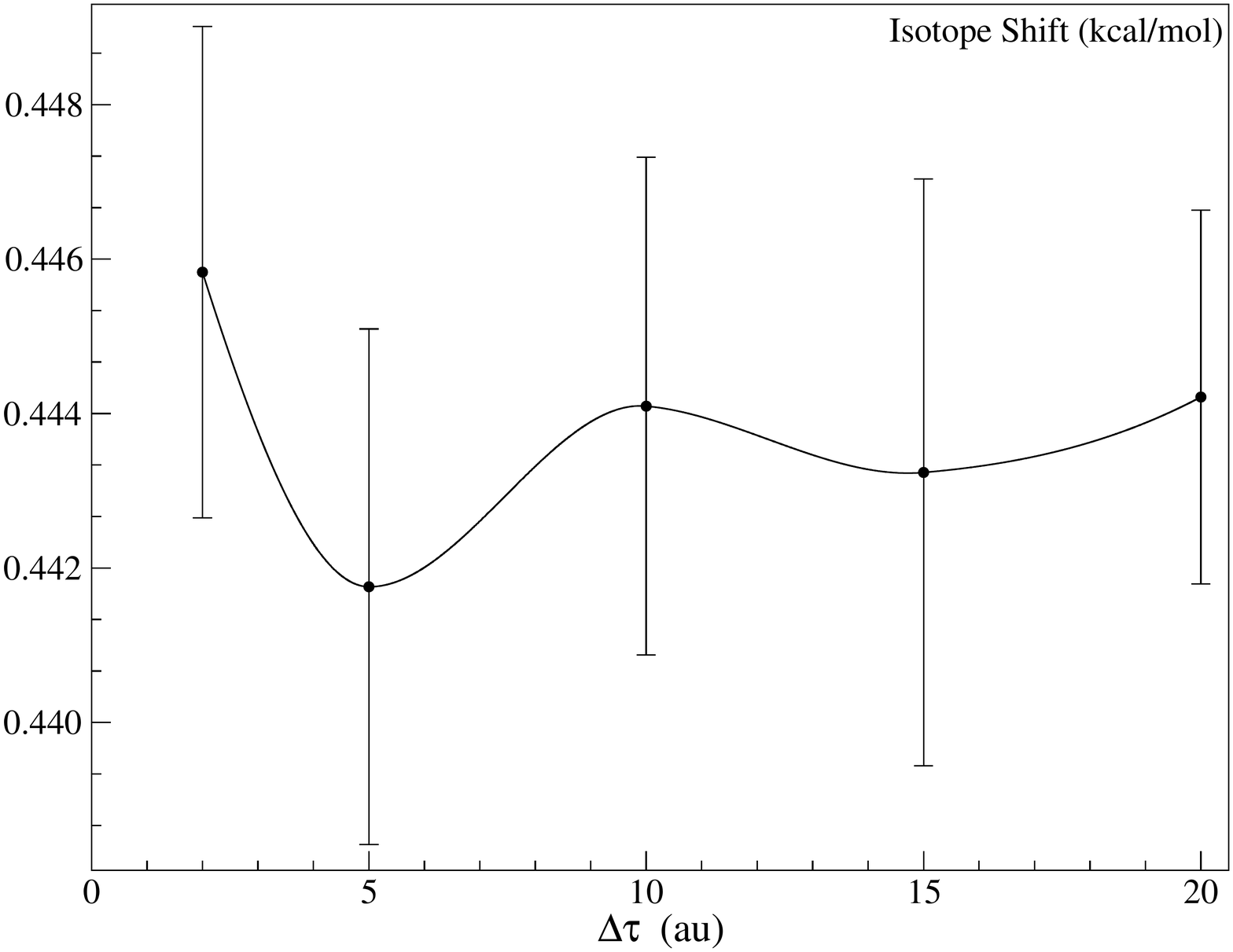}
\caption{DMC energies for the water dimer and monomer with $N_w =1.96 \times 10^4$ walkers for five values of the time step $\Delta\tau$. Time step values 
range from $2.0$ to $20.0$ au. All calculations were run for a total projection time of $2.0 \times 10^6$ au. Top Left: H$_2$O dimer energies. Top Right: 
H$_2$O monomer energies. Middle Left: D$_2$O dimer energies. Middle Right: D$_2$O monomer energies. Bottom Left: Binding energy $D_0$ for H$_2$O and D$_2$O 
dimers. Bottom Right: Isotope shift $\delta D_0$ for the dimer binding energies (note the scale). The ``classical'' energy (here and below) is computed by using the values of the corresponding potential energy minima.}
\label{fig:timestep}
\end{figure*}

\begin{figure*}
\includegraphics[width=0.495\textwidth]{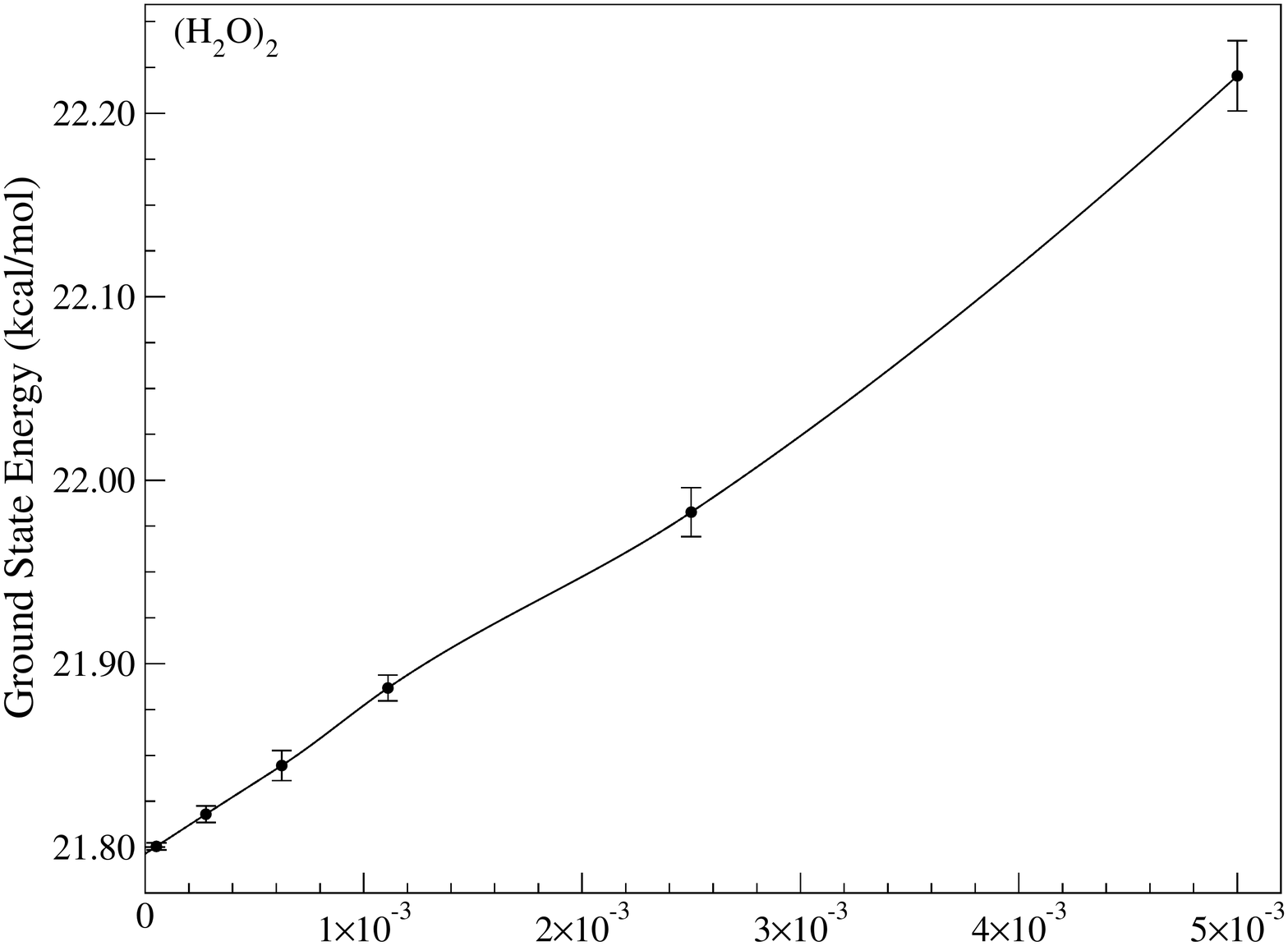}
\includegraphics[width=0.495\textwidth]{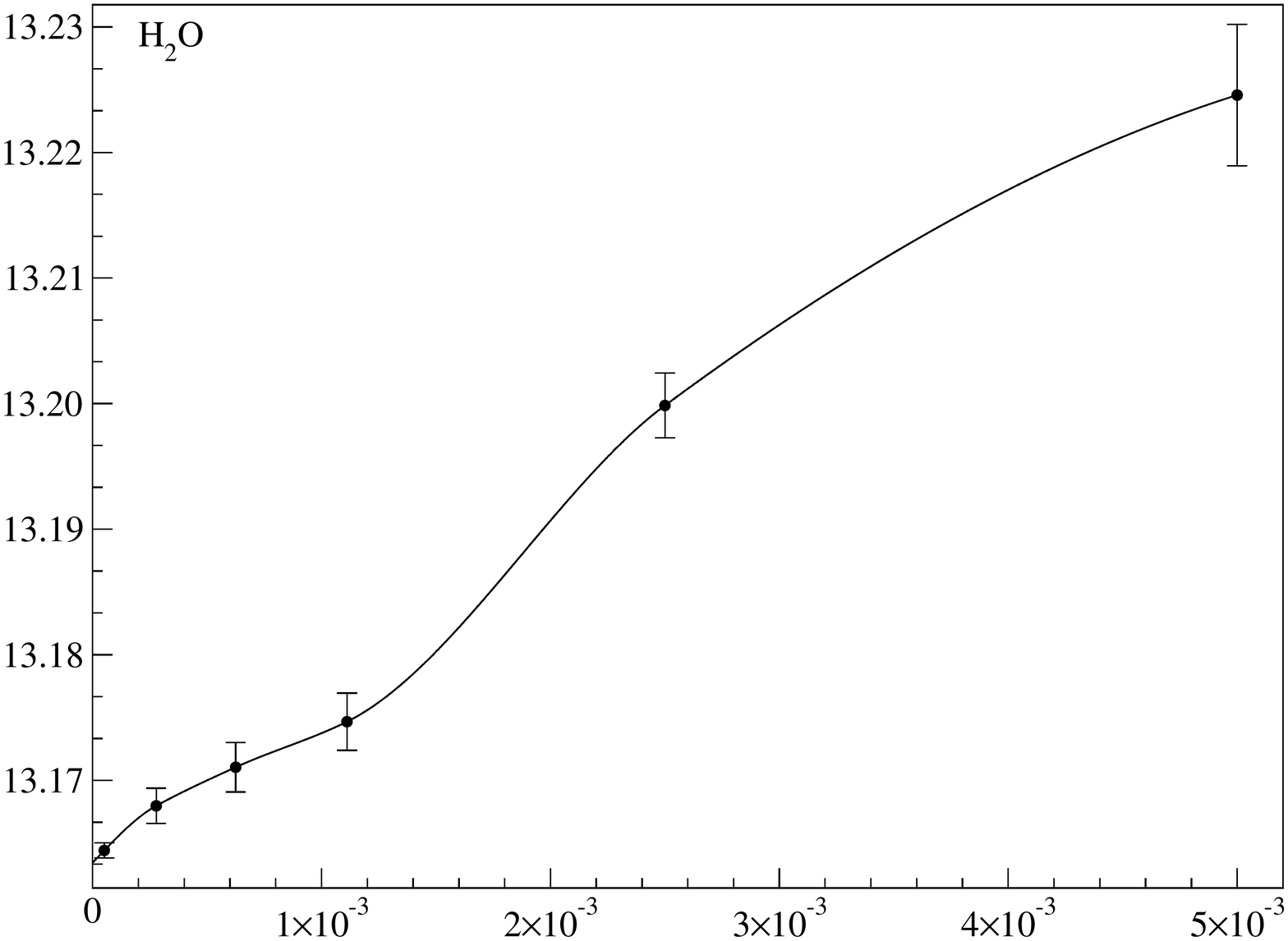}
\includegraphics[width=0.495\textwidth]{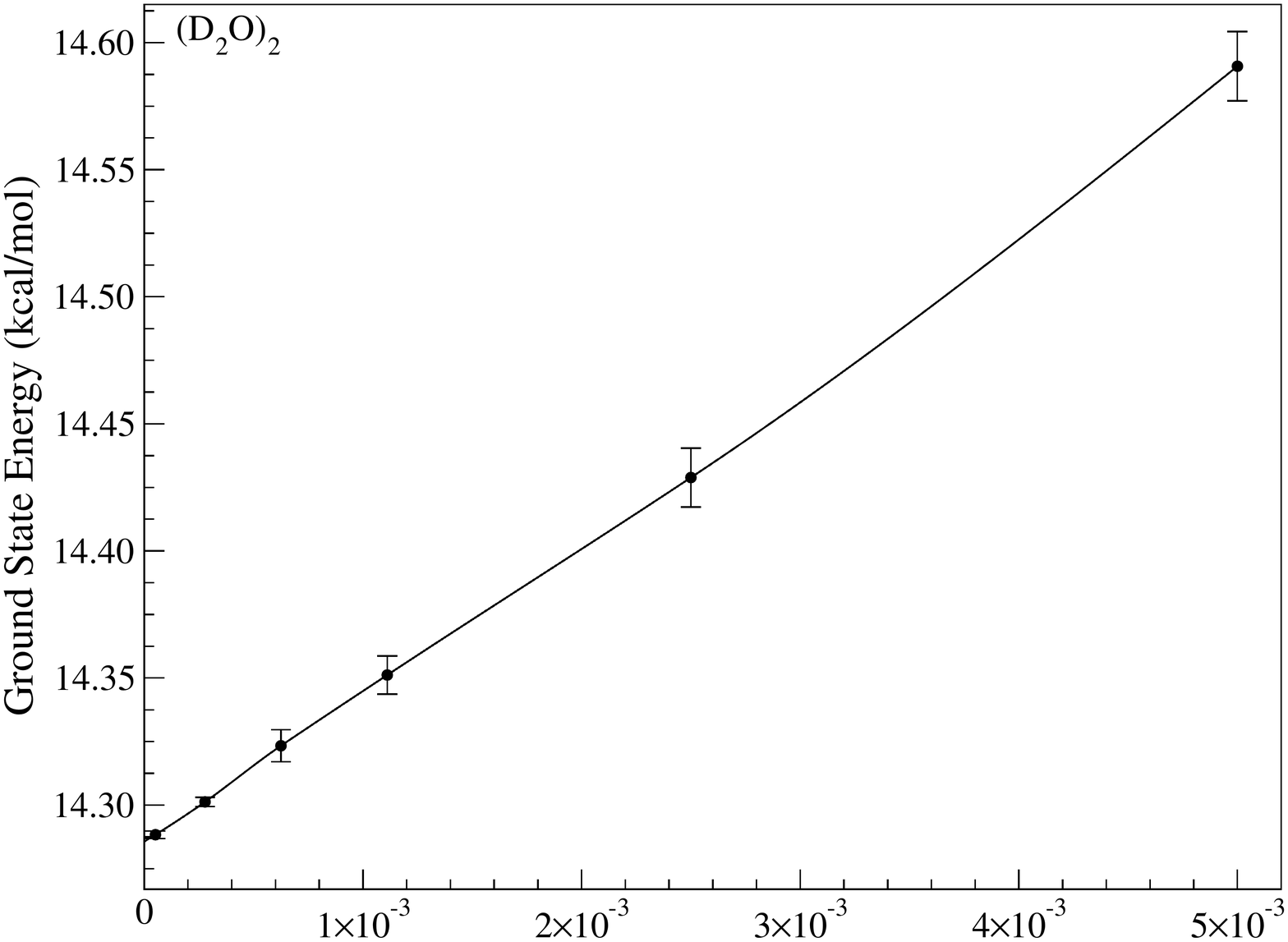}
\includegraphics[width=0.495\textwidth]{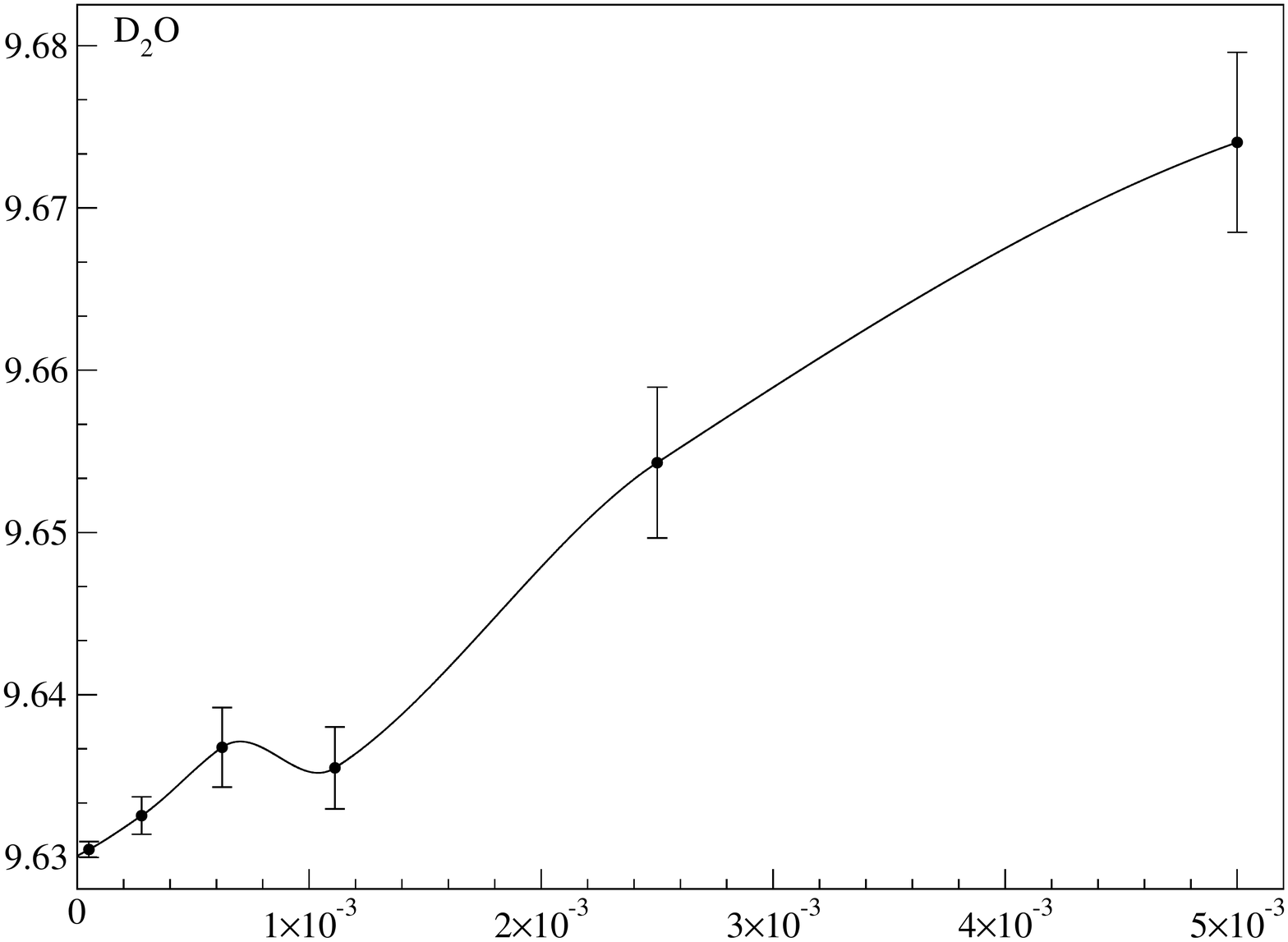}
\includegraphics[width=0.495\textwidth]{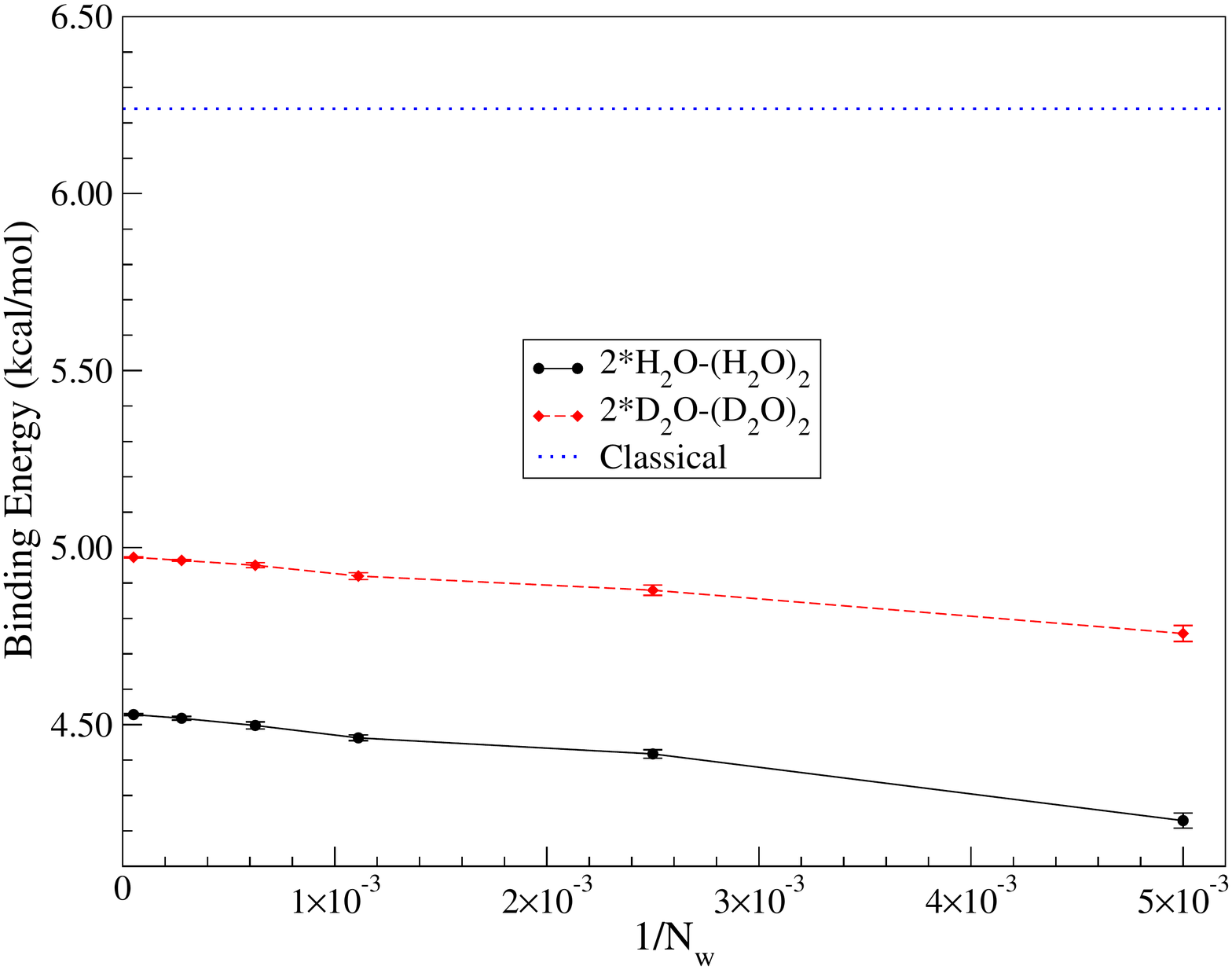}
\includegraphics[width=0.495\textwidth]{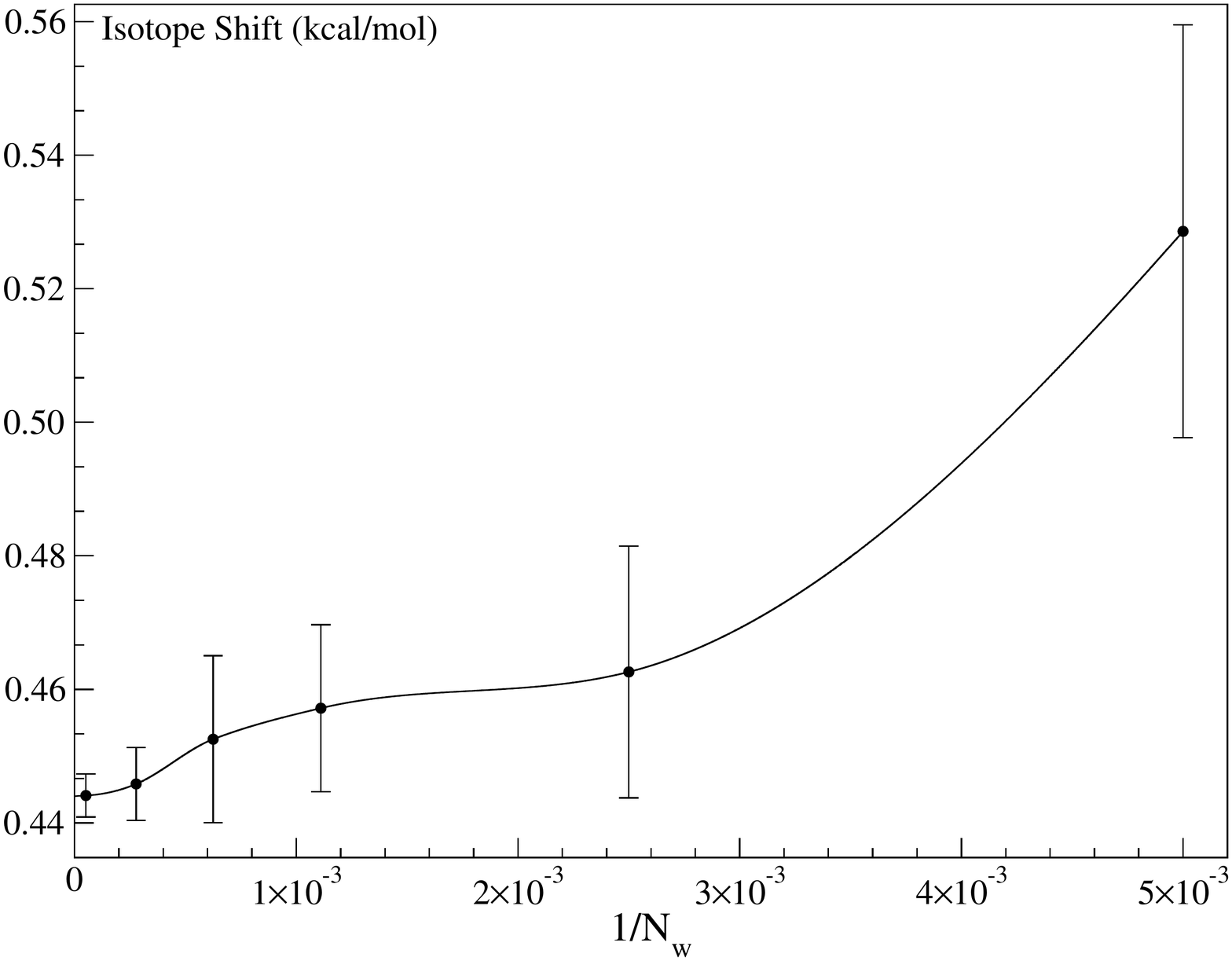}
\caption{Same as \Fig{fig:timestep} but as a function of the reciprocal walker number $1/N_w $ with fixed value of the time step $\Delta\tau=10.0$ au. Walker numbers range from 
$2.0 \times 10^2$ to $1.96 \times 10^4$.}
\label{fig:recipN}
\end{figure*}

Clearly, the total numerical cost scales as
\be
\label{eq:numcost}
\mbox{numerical cost} \sim \tau_{\rm max} N_w 
\ee
and for given $\Delta\tau$ and $N_w$ the statistical error should scale as $\sim\tau_{\rm max}^{-1/2}$  Because the total projection time is fixed, it follows from \Eq{eq:numcost} that 
for a constant value of $\Delta\tau$, the numerical effort should grow linearly with population size $N_w$. It is expected that for a fixed total projection time the statistical error 
should scale as $\sim N_w^{-1/2}$. This is demonstrated in \Fig{fig:errorbars} using the DMC results for the water monomer and dimer with $\tau_{\rm max}=2.0 \times 10^6$ au. Therefore,
the statistical error should scale as 
\be
\mbox{statistical error}\sim\left(N_w\tau_{\rm max}\right)^{-1/2}\sim (\mbox{numerical cost})^{-1/2}
\ee
That is, the statistical error is a function of the numerical cost only. Consequently, we assert that no computational time is gained through the use of small walker numbers, as for simulations with 
fewer random walkers, the projection time must be increased accordingly to maintain the same statistical error. At the same time, the systematic bias in $N_w$ is reduced with increasing population size.
This analysis suggests that in order to reduce the statistical error it is preferable to consider larger population sizes $N_w$ rather than longer simulation times $\tau_{\rm max}$. 
Of course, one should also take into account other considerations, such as limitations in the computer memory, and also the need to perform a sufficiently long equilibration, during which the averages are 
not accumulated.

\Fig{fig:timestep} and \Fig{fig:recipN} show the dependence of the DMC energy estimates on the time step $\Delta\tau$, and the reciprocal walker number $1/N_w$, respectively. We observe that the 
time step bias for the ground state energy estimate  at $\Delta\tau=10.0$ au is noticeable. For example, for the water dimer it is $\sim 0.03$ kcal/mol. However, the estimate of the binding energy,
\be
D_0:=2E_{\rm H_2O}-E_{\rm (H_2O)_2}
\ee
is hardly sensitive to the time step due to the nearly complete cancellation of the systematic errors (see the bottom left panel of \Fig{fig:timestep}). At the same time, the population size bias for the binding 
energy estimate (the bottom left panel of \Fig{fig:recipN}) is more evident and does not disappear completely. This behavior is attributed to the fact that the population of random walkers represents 
the wavefunction and directly reflects the geometric composition of the system. For example, the population size bias is consistently stronger for the dimer than the monomer. In other words, the DMC energy 
estimate converges faster for the monomer than the dimer, which results in imperfect compensation of the systematic errors. Therefore, the failure of the biases for both systems to fully cancel appears in 
the binding energy curves as a residual bias (bottom left panel of \Fig{fig:recipN}). Yet, for the isotope shift (the lowest right panel of \Fig{fig:recipN}), 
\be
\delta D_0:=D_0({\rm  D_2O})-D_0({\rm  H_2O})
\ee
the population size bias is significantly less pronounced. The systematic errors do cancel in this case, and the isotope shift converges to a value near $\delta D_0 =0.444$ kcal/mol. This weak 
dependence of the isotope shift on walker population indicates that the extent and behavior of the binding energy biases are similar for both types of isotopomers.

Our results extrapolated to $N_w \to \infty$ are summarized in Table~\ref{tab:ded0}. Estimates of the statistical uncertainty in the DMC energies are at least an order of magnitude smaller than the 
systematic error ($0.001$ kcal/mol or lower) for the largest $N_w$ considered in Figs.~\ref{fig:recipN} and~\ref{fig:hexamer} and, as such, are not included in the table. 
By extrapolating the absolute ground state energy estimates $E_0$ to the $\Delta\tau \to 0$ limit we could significantly improve their estimates. For example, for the water monomer H$_2$O we would obtain $E_0=13.18$ kcal/mol, which coincides with the numerically exact value obtained by diagonalizing the Hamiltonian using a discrete variable representation. However, we emphasize that such extrapolation prior to taking the energy difference would be a bad idea, as in this case we would not be able to take advantage of the systematic error cancellations. 
Thus, while a bias does indeed exist for $E_0$, we determined that the time step error in $D_0$ is negligibly small for the dimer and is predicted to be so for the hexamer even with $\Delta\tau=10$ au.

We find that our DMC estimates of $D_0$ for both (H$_2$O)$_2$ and (D$_2$O)$_2$ are about 1.5 kcal/mol higher than the experimental values reported in refs. \onlinecite{reisler2011,reisler2012}, which is a clear indication of the failure of the q-TIP4P/F PES in describing accurately the energetics of small water clusters. Interestingly though, for the isotope shift in the binding energy the present result ($\sim0.44$ kcal/mol) happened to agree well with the experiment ($\sim0.40$ kcal/mol).

\begin{table}
\caption{Classical ($D_e$) and quantum ($D_0$) estimates of the binding energies for the water clusters considered in this work. $E_0$ stands for the ground state energy estimates. 
All the energy estimates are obtained by extrapolating the DMC results with $\Delta\tau=10.0$ au to the $N_w \to \infty$ limit. Experimental binding energies for the dimer $D_0^{\rm expt.}$ 
were obtained from refs.~\onlinecite{reisler2011,reisler2012}. All energies are reported in kcal/mol.}
\label{tab:ded0}
\begin{tabular}{c|ccccc}
\hline\hline
Structure & $E_{\rm min}$ & $E_0$ & $D_e$ & $D_0$ & $D_0^{\rm expt.}$ \\
\hline
H$_2$O & 0.00 & 13.16 & - & - & - \\
(H$_2$O)$_2$ & -6.24 & 21.80 & 6.24 & 4.53 & $3.16 \pm 0.03$ \\
(H$_2$O)$_6$-Cage & -50.64 & 41.09 & 50.64 & 37.89 & - \\
(H$_2$O)$_6$-Prism & -50.19 & 41.73 & 50.19 & 37.24 & - \\
\hline
D$_2$O & 0.00 & 9.63 & - & - & - \\
(D$_2$O)$_2$ & -6.24 & 14.29 & 6.24 & 4.97 & $3.56 \pm 0.03$ \\
(D$_2$O)$_6$-Cage & -50.64 & 17.09 & 50.64 & 40.61 & -\\
(D$_2$O)$_6$-Prism & -50.19 & 17.69 & 50.19 & 40.05 & - \\
\hline\hline
\end{tabular}
\end{table}

\section*{Water Hexamer: Cage versus Prism.}
\label{sect:hexamer}
This section reports the results of DMC simulations for the water hexamer and its (D$_2$O)$_6$ isotopomer. Specifically, we have performed ground state calculations for two low-lying isomers: 
the cage and prism. The cage geometry corresponds to the global minimum of the q-TIP4P/F PES, while the prism local minimum is only 0.45 kcal/mol higher.
Originally, we also planned to consider the next-highest isomer in the sequence, the book, which lies about 1.2 kcal/mol higher than the cage minimum, but discovered that it is extremely unstable due to very low energy barrier separating it from the prism minimum. For example, in a classical Monte Carlo simulation 
a random walk starting in the book configuration would quickly jump into the prism minimum at temperatures as low as $T\sim 5$K, thus making the book isomer physically undetectable in a hypothetical experiment. Moreover, 
the quantum book isomer has even less physical meaning because including the quantum effects would only destabilize the system to a greater extent. In contrast to the book structure, the 
classical prism isomer appears to be stable up to temperatures as high as $T\sim 60$K, which makes the q-TIP4P/F prism geometry much less ambiguous to define and much more likely to be observed in a hypothetical experiment. 
However, the existence of relatively high potential energy barriers surrounding the prism isomer does not necessarily prevent the random walkers in the DMC simulation from leaking into other 
local minima, which happens often enough to make the problem nontrivial both conceptually and numerically. It is sufficient for only one random walker to escape into another energy minimum, 
where it may start replicating, thereby quickly causing a significant portion of the random walker population to represent the ``wrong'' geometry. 
As mentioned above, this problem can be circumvented in principle by using importance sampling (see, e.g. refs.~\onlinecite{assaraf2000,watts1991,umrigar1993,warren2006}).
However, 
these approaches may not be feasible to implement for systems as complex as water clusters, when an appropriate trial wavefunction must be very nontrivial to both determine and parametrize. Therefore, in the present 
context, we find the use of geometric constraints less problematic. Intelligently chosen geometric constraints impose artificial barriers on the random walkers, preventing them from escaping out 
of the region defined by the particular isomer. 
Furthermore, the problem exists even in calculations of the true ground state, i.e., regardless of the relationship between the latter and the isomer in question. In such a calculation the random walkers 
are initialized, usually but not always, in the global energy minimum (for  the q-TIP4P/F PES, it is the cage minimum). Occasionally, one of the random walkers may still jump from the cage minimum to one of the local minima, 
where it has a chance to be replicated. Arguably, due to either physical or unphysical reasons, this migration of random walkers leads to their population being delocalized over energy minima representing different isomers. Consequently, for the present study, we find it necessary to impose geometric constraints even for DMC calculations involving the true ground state.

\begin{figure}
 \caption{The pair distance order parameter $\Delta\rho_{\rm pd}$ ({\it cf.} \Eq{eq:pd}) shown for several randomly chosen walkers as a function of projection time for two typical DMC simulations using $N_w=8\times 10^5$ with all the random walkers initialized in the prism minimum.  Top: An unconstrained calculation in which the walkers started to leak  
out of the prism minimum into different regions of the PES at $\tau\sim 10^5$ au. Bottom: Same, except that geometric constraints were imposed to prevent walker leakage.} 
\label{fig:constraint}
\includegraphics[width=0.495\textwidth]{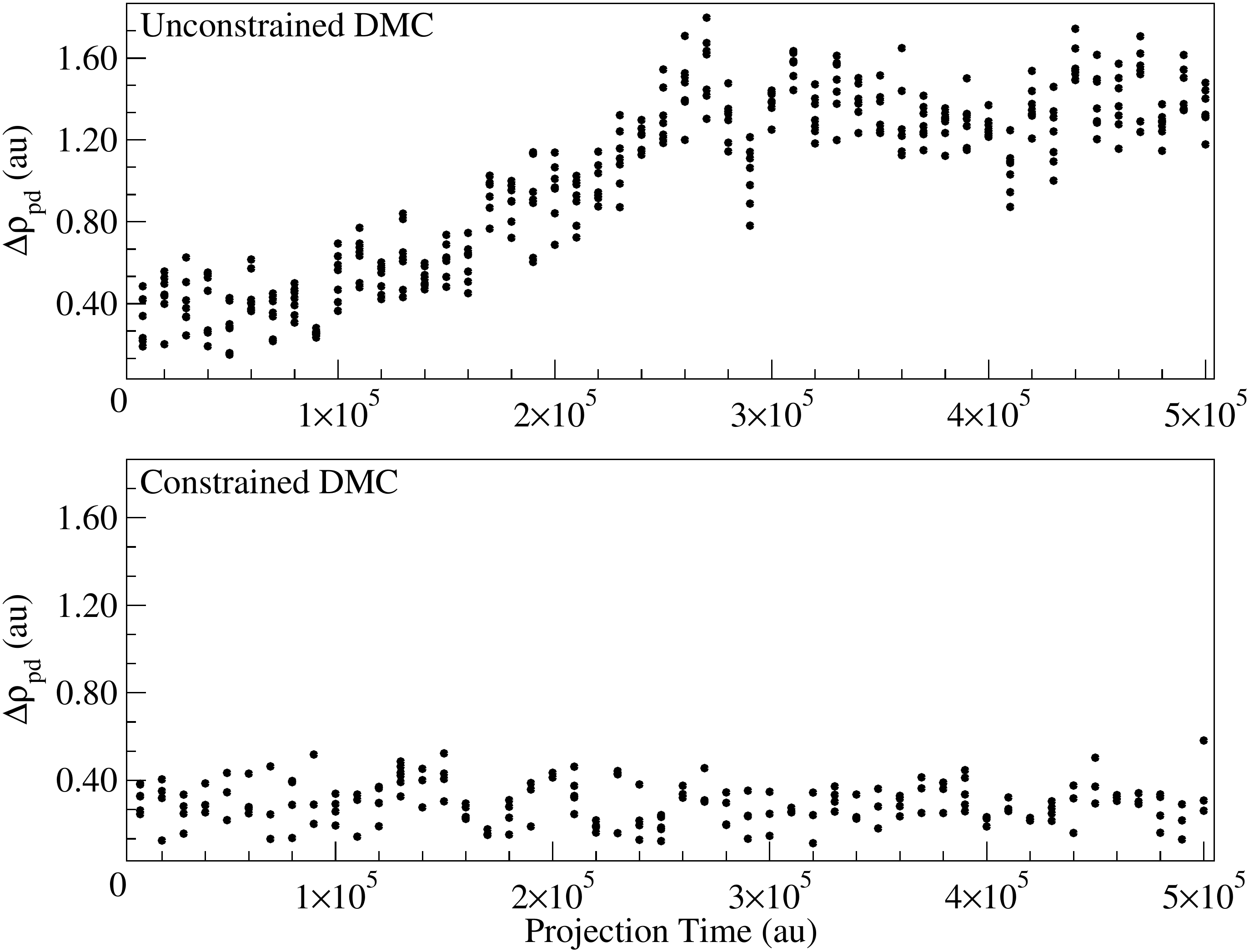}
\end{figure}

The isomers of water clusters are usually identified by the arrangement of the oxygen atoms only; there might be several local minima, different only by the orientation of one or two water 
molecules, which are still identified by the same geometric motif (e.g. the prism). These minima are nearly isoenergetic, and they are often separated by low potential energy barriers. To 
this end, consider a configuration $\br=(\br_1,...,\br_K)$ ($\br_i \in\mathbb{R}^3$) of $K$ oxygen atoms (here $K=6$) and a vector $\brho_{\rm pd}[\br] \in\mathbb{R}^{K(K-1)/2}$ with elements defined by all the $K(K-1)/2$ 
pair distances $|\br_i-\br_j|$ ($i>j$) arranged in descending order. The following constraint can then be deployed to prevent the random walkers from leaving the basin of attraction defined by 
a reference configuration $\br^{\rm (ref)}$ (e.g., corresponding to the prism minimum of the PES):
\be\label{eq:pd}
\Delta\rho_{\rm pd}:=\frac{\sqrt{2}}{\sqrt{K(K-1)}}\left|\brho_{\rm pd}[\br]-\brho_{\rm pd}[\br^{\rm (ref)}] \right|< h_{\rm pd}
\ee
where the full expression is the root-mean-square displacement of the instantaneous pair distances from those of the initial reference structure, and $h_{\rm pd}$ is an empirically chosen parameter. 
Note that the pair distance constraint \eqref{eq:pd} has two important properties: it is rotationally and translationally invariant and, secondly, its numerical evaluation is inexpensive, so it 
can be implemented at every MC iteration without a significant increase in the overall computational cost. However, due to its simplicity, it may not always be very effective. 

Consequently, we implemented two additional geometric constraints, which follow from reasoning analogous to that used in formulating \Eq{eq:pd}. The second constraint is based on the metric defined 
by the three principal moments of inertia arranged in descending order within the vector $\brho_{\rm mi}[\br]\in\mathbb{R}^{3}$: 
\be\label{eq:mi}
\Delta\rho_{\rm mi}:=\frac{1}{\sqrt{3}}\left|\brho_{\rm mi}[\br]-\brho_{\rm mi}[\br^{\rm (ref)}] \right|< h_{\rm mi}
\ee
Similarly, the vector $\brho_{\rm cm}[\br] \in\mathbb{R}^{K}$ with elements $\left |\br_j-\br_{\rm cm} \right|$ arranged in descending order contains the distances between all $K$ oxygen atoms and the 
center of mass $\br_{\rm cm}$ of the water cluster. Thus, the third constraint is defined by 
\be\label{eq:cm}
\Delta\rho_{\rm cm}:=\frac{1}{\sqrt{K}}\left|\brho_{\rm cm}[\br]-\brho_{\rm cm}[\br^{\rm (ref)}] \right|< h_{\rm cm}
\ee
The threshold values $h_{\rm pd}$, $h_{\rm mi}$, and $h_{\rm cm}$ are established empirically by trial and error. For example, we discovered that the cage ground state calculation requires only the pair 
distances as a constraint with $h_{\rm pd}=0.76$ au, while for the prism ground state calculation all three constraints are necessary with $h_{\rm pd}=0.76$ au, $h_{\rm mi}=125$ amu~${\rm au^2}$, and 
$h_{\rm cm}=0.66$ au. 

\Fig{fig:constraint} shows an example from two typical DMC simulations, without geometric constraints (top) and with all three constraints implemented using the prism as a reference configuration. The quantity shown is that defined by \Eq{eq:pd} for several randomly chosen walkers as a function of projection time $\tau$. The leaking out of the prism local minimum in the unconstrained simulation starts to occur at $\tau\sim10^5$ au.

\begin{figure*}
\includegraphics[width=0.495\textwidth]{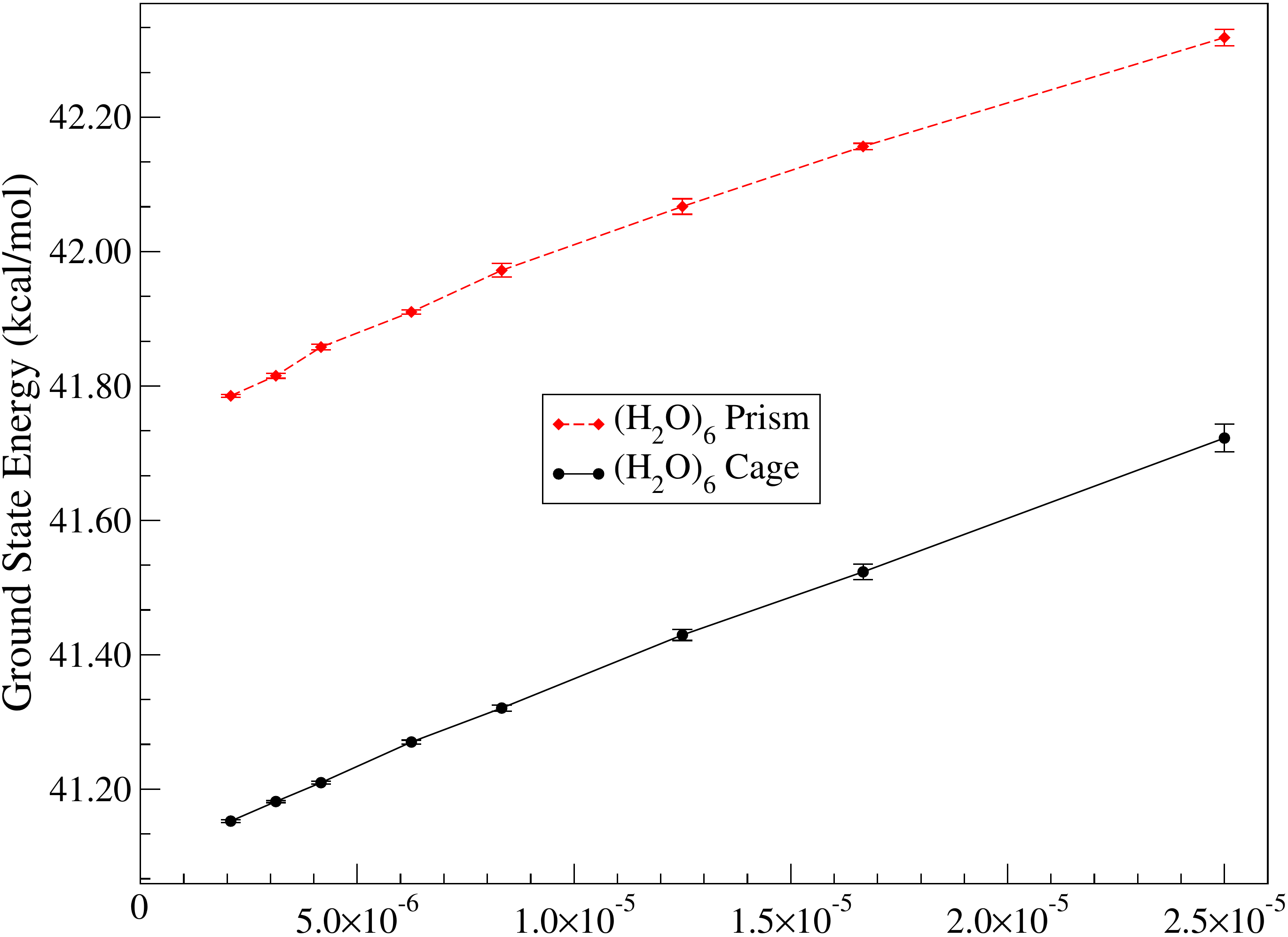}
\includegraphics[width=0.495\textwidth]{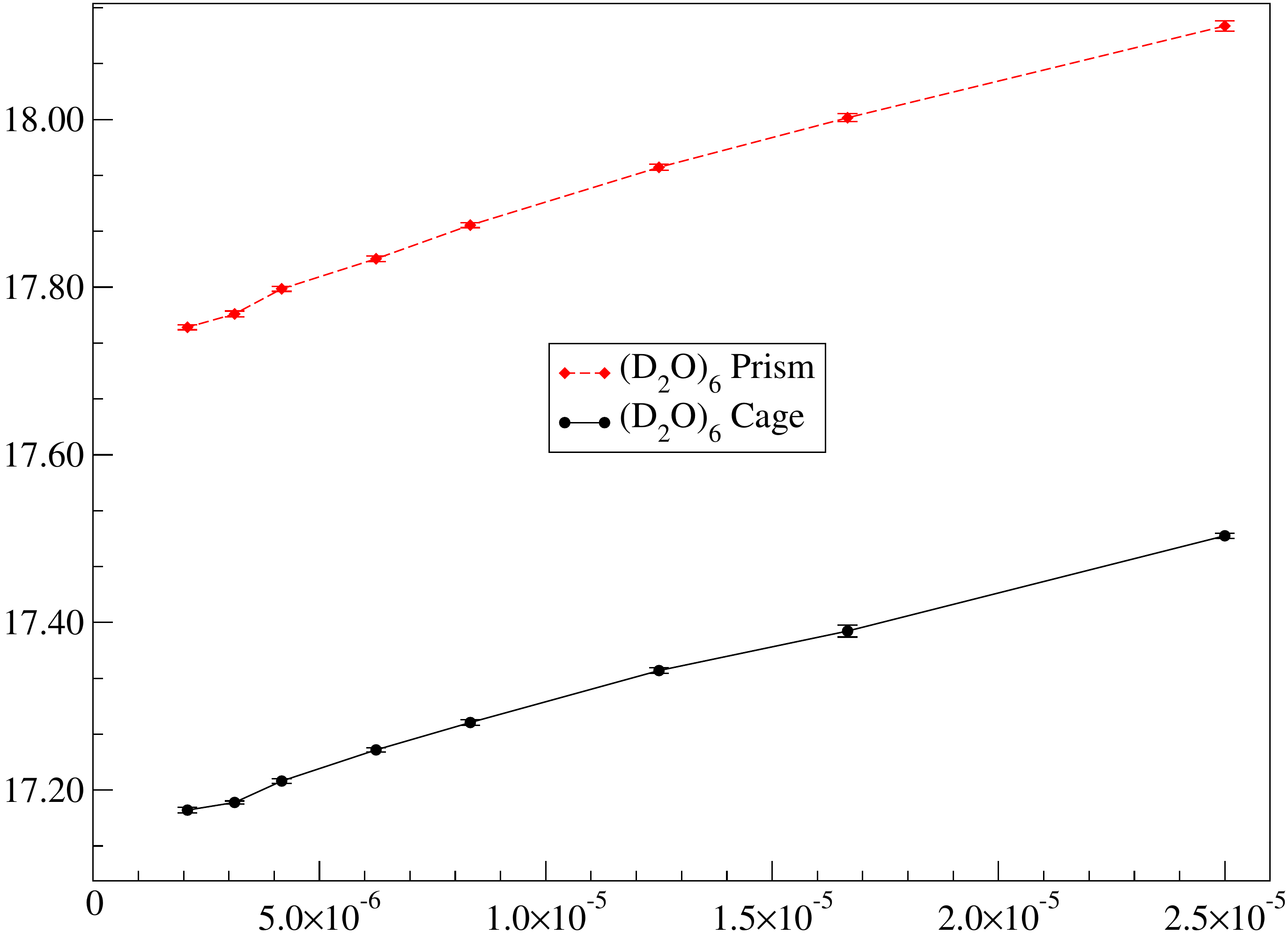}
\includegraphics[width=0.495\textwidth]{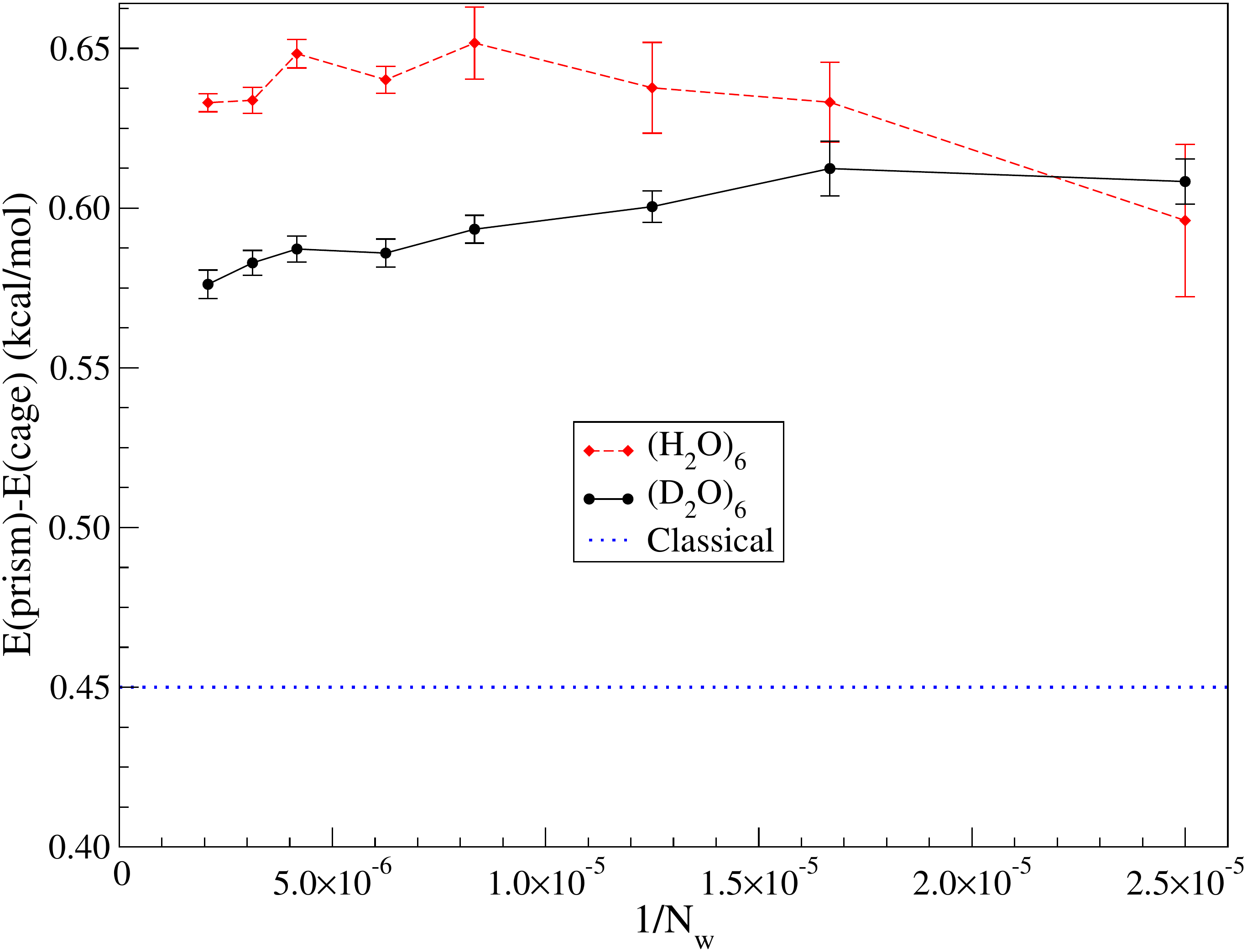}
\includegraphics[width=0.495\textwidth]{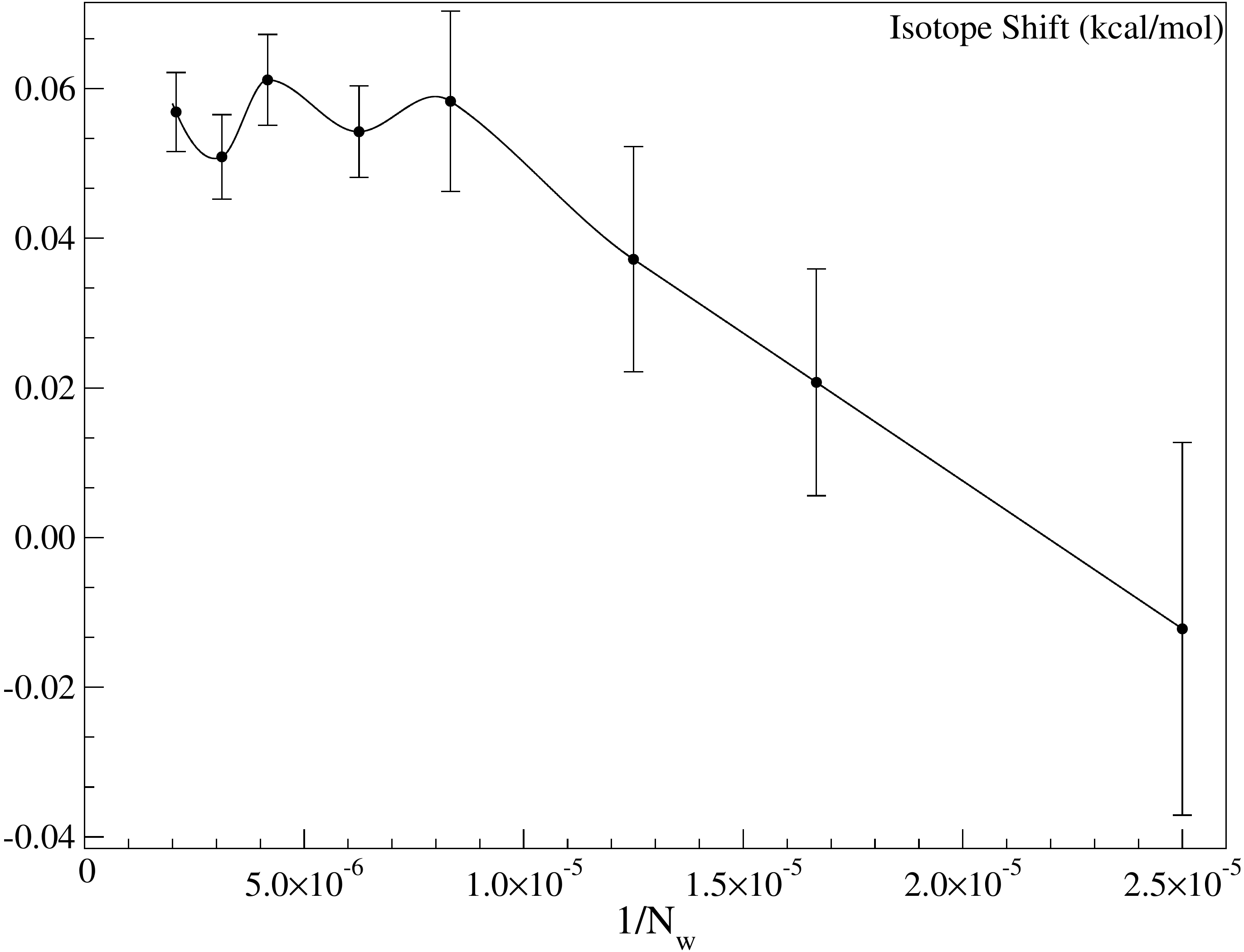}
\caption{DMC energies for the water hexamer with $\Delta\tau=10$ au as a function of the reciprocal walker number $1/N_w $. The walker numbers used in the simulations range from 
$N_w=4.0\times10^4$ to $4.8\times10^5$. Top Left: Ground state energies $E_0$ for (H$_2$O)$_6$ cage and prism. Top Right: Same, for (D$_2$O)$_6$ cage and prism. Bottom Left: Prism-cage 
energy differences for (H$_2$O)$_6$ and (D$_2$O)$_6$ isotopomers $\Delta E= E_0({\rm prism})-E_0({\rm cage})$. Bottom Right: Isotope shift for prism-cage energy differences 
$\delta_{\rm isotope} = \Delta E_{\rm H_2O}-\Delta E_{\rm D_2O}$ (note the scale).}
\label{fig:hexamer}
\end{figure*}

\Fig{fig:hexamer} shows the DMC results for H$_2$O hexamer prism and cage and their D$_2$O isotopomers with fixed $\Delta\tau=10$ au and varying population size: $N_w=4.0\times10^4$, 
$6.0\times10^4$, $8.0\times10^4$, $1.2\times10^5$, $1.6\times10^5$, $2.4\times10^5$, $3.2\times10^5$, and $4.8\times10^5$. The total projection time for most simulations was on the order 
of $\tau_{\rm max}\sim 10^6$, which, as established earlier for the dimer, means that simulations with larger $N_w$ have smaller statistical errors. An examination 
of these plots reveals that for the population sizes considered, the bias in $N_w$ for the ground state energies of the hexamer is noticeable. For example, the change of the DMC energy 
estimate for the cage isomer is about 0.03 kcal/mol when the population size is increased from $N_w=3.2\times10^5$ to $4.8\times10^5$. Moreover, for $N_w= 4.8\times10^5$, which is the largest 
walker number used in this work, the energy estimate appears to be about 0.06 kcal/mol higher than the value extrapolated to the $N_w\to\infty$ limit. At the same time, \Fig{fig:hexamer} 
shows that the biases in the absolute prism and cage energies follow a nearly identical trend regardless of which isotopomer is considered. Energies for (H$_2$O)$_6$ appear to be slightly 
less converged than those for (D$_2$O)$_6$ at the same number of walkers and projection times (i.e., statistical errors are smaller for the ``less quantum'' (D$_2$O)$_6$). 
The slopes of the prism and cage energies as a function of $1/N_w$ are virtually equivalent (even up to small features), which indicates that the bias is not dramatically affected by 
variations in the spatial arrangement of the monomers. As a consequence, upon taking the energy difference between the prism and cage isomers, the bias is substantially reduced resulting only in a 
weak dependence on $N_w$. Visual inspection of the energy curves enables us to conclude that our DMC estimates of the ground state energy differences, 
\be\label{eq:DE}
\Delta E= E_0({\rm prism})-E_0({\rm cage})
\ee
are accurate to about 0.01 kcal/mol, when sufficiently large walker numbers (i.e., $N_w \ge 2.0\times10^5$) are considered. Additionally, the small oscillations in $1/N_w $ that still 
remain in the DMC energy estimates correlate between (H$_2$O)$_6$ and (D$_2$O)$_6$, which makes us believe that these oscillations are real.

Considering now the isotope shift (as shown in the lower left panel of \Fig{fig:hexamer}),
\be\label{eq:dE}
\delta_{\rm isotope}:= \Delta E_{\rm H_2O}-\Delta E_{\rm D_2O}
\ee
we conclude that further cancellations of the systematic errors occur for this quantity, such that the result seems to reach a plateau near $\delta_{\rm isotope} = 0.055\pm 0.01$ kcal/mol at 
$N_w=1.2\times10^5$. However, compared to this small value of the isotope shift, the bias for walker numbers $N_w<1.2\times10^5$ is noticeable, and at $N_w\sim5.0\times10^4$  
$\delta_{\rm isotope}$ even changes sign.

\section*{Conclusions}
\label{sect:conclusion}
We have provided an assessment of the DMC method of Anderson.  In particular, we have analyzed the bias with respect to the time step $\Delta\tau$ and the walker number $N_w$ through application 
to the water monomer, dimer, and hexamer along with their D$_2$O isotopomers. We have determined that the use of relatively small walker numbers $N_w$ is unprofitable because longer projection 
times $\tau_{\rm max}$ are then required to maintain the same statistical error. Consequently, the best-case scenario is to choose larger values of $N_w$, as both the statistical and systematic 
errors decrease with increasing $N_w$.

The calculations undertaken in this work for the hexamer used walker numbers up to $N_w=4.8\times10^5$ with total projection times on the order of $\tau_{\rm max}\sim 10^6$, which for $\Delta\tau=10$ au 
required $~\sim 10^5\times N_w$ potential energy evaluations per isomer. Nevertheless, the absolute ground state energies did not level off to a well-defined plateau. As demonstrated here, obtaining 
accurate estimates for the binding energies does require high accuracy for the absolute values of the ground state energies. Indeed, the population size bias for the water dimer does not cancel with 
that of the water monomer, for which the bias is negligible. However, this is not an issue for the time step bias, as the systematic errors for the dimer and monomer are removed almost completely upon 
determination of the binding energy. At the same time, the biases undergo cancellation for interesting properties of the water hexamer such as the prism-cage energy difference and isotope shift. 
Therefore, it is possible to obtain accurate values for these quantities in practice. However, without carrying out a systematic study, as done in this work, one cannot assume that the errors would 
fortuitously cancel. Also, the q-TIP4P/F PES renders our calculations less expensive by orders of magnitude compared to presumably more accurate {\it ab initio}-based 
surfaces\cite{wang2011,medders2013,babin2014} or even true {\it ab initio} surfaces. Should we use one of these surfaces, this publication in 2015 would hardly be possible. 

Interestingly, the quantum effect for the cage-prism energy difference, $\delta_{\rm QM}:= \Delta E_{\rm H_2O}-\Delta E_{\rm classical}=0.20$ kcal/mol, is relatively small, while the isotope shift 
$\delta_{\rm isotope}=0.055$ kcal/mol is even smaller. Both these results are consistent with an earlier prediction in which the Self Consistent Phonons method 
was applied to the water hexamer using several different water models\cite{brown2014}. In light of this conclusion, we cannot ignore the DMC result of ref. \onlinecite{wang2012} regarding the 
water hexamer (albeit using the WHBB PES of ref. \onlinecite{wang2011}), which indicated that the cage and prism bound state energies are nearly the same, 
$E_0({\rm prism})\approx E_0({\rm cage})$, while the classical cage-prism energy difference for this PES is $\Delta E_{\rm classical}\approx 0.4$ kcal/mol. The maximum number of random walkers $N_w=1.6\times10^5$ and the total projection time $\tau_{\rm max}=8.0\times 10^5$ au used in ref. \onlinecite{wang2012} give rise to statistical errors noticeably larger than those in the present study. 
However, the most important difference between the present DMC study and that of ref. \onlinecite{wang2012} is that no geometric constraints were implemented in the latter, while we found the geometric 
constraints to be of crucial importance in preventing the random walker population from spilling out of the potential energy minimum corresponding to a particular isomer. If such a procedure is not 
implemented, one should keep his or her fingers crossed in the hope that during the course of the DMC simulation the walker population does not spread over several local minima (corresponding to the ``wrong'' isomers), thereby resulting in an incorrect energy estimate.

\section*{Acknowledgements}
This work was supported by the National Science Foundation (NSF) Grant No. \mbox{CHE-1152845}.
We thank Anne McCoy for useful discussions of the DMC method.


\begin{thebibliography}{36}%
\makeatletter
\providecommand \@ifxundefined [1]{%
 \@ifx{#1\undefined}
}%
\providecommand \@ifnum [1]{%
 \ifnum #1\expandafter \@firstoftwo
 \else \expandafter \@secondoftwo
 \fi
}%
\providecommand \@ifx [1]{%
 \ifx #1\expandafter \@firstoftwo
 \else \expandafter \@secondoftwo
 \fi
}%
\providecommand \natexlab [1]{#1}%
\providecommand \enquote  [1]{``#1''}%
\providecommand \bibnamefont  [1]{#1}%
\providecommand \bibfnamefont [1]{#1}%
\providecommand \citenamefont [1]{#1}%
\providecommand \href@noop [0]{\@secondoftwo}%
\providecommand \href [0]{\begingroup \@sanitize@url \@href}%
\providecommand \@href[1]{\@@startlink{#1}\@@href}%
\providecommand \@@href[1]{\endgroup#1\@@endlink}%
\providecommand \@sanitize@url [0]{\catcode `\\12\catcode `\$12\catcode
  `\&12\catcode `\#12\catcode `\^12\catcode `\_12\catcode `\%12\relax}%
\providecommand \@@startlink[1]{}%
\providecommand \@@endlink[0]{}%
\providecommand \url  [0]{\begingroup\@sanitize@url \@url }%
\providecommand \@url [1]{\endgroup\@href {#1}{\urlprefix }}%
\providecommand \urlprefix  [0]{URL }%
\providecommand \Eprint [0]{\href }%
\providecommand \doibase [0]{http://dx.doi.org/}%
\providecommand \selectlanguage [0]{\@gobble}%
\providecommand \bibinfo  [0]{\@secondoftwo}%
\providecommand \bibfield  [0]{\@secondoftwo}%
\providecommand \translation [1]{[#1]}%
\providecommand \BibitemOpen [0]{}%
\providecommand \bibitemStop [0]{}%
\providecommand \bibitemNoStop [0]{.\EOS\space}%
\providecommand \EOS [0]{\spacefactor3000\relax}%
\providecommand \BibitemShut  [1]{\csname bibitem#1\endcsname}%
\let\auto@bib@innerbib\@empty
\bibitem [{\citenamefont {Anderson}(1975)}]{anderson1975}%
  \BibitemOpen
  \bibfield  {author} {\bibinfo {author} {\bibfnamefont {J.~B.}\ \bibnamefont
  {Anderson}},\ }\href@noop {} {\bibfield  {journal} {\bibinfo  {journal} {J.
  Chem. Phys.}\ }\textbf {\bibinfo {volume} {63}},\ \bibinfo {pages} {1499}
  (\bibinfo {year} {1975})}\BibitemShut {NoStop}%
\bibitem [{\citenamefont {Anderson}(1976)}]{anderson1976}%
  \BibitemOpen
  \bibfield  {author} {\bibinfo {author} {\bibfnamefont {J.~B.}\ \bibnamefont
  {Anderson}},\ }\href@noop {} {\bibfield  {journal} {\bibinfo  {journal} {J.
  Chem. Phys.}\ }\textbf {\bibinfo {volume} {65}},\ \bibinfo {pages} {4121}
  (\bibinfo {year} {1976})}\BibitemShut {NoStop}%
\bibitem [{\citenamefont {Viel}\ and\ \citenamefont {Whaley}(2001)}]{viel2001}%
  \BibitemOpen
  \bibfield  {author} {\bibinfo {author} {\bibfnamefont {A.}~\bibnamefont
  {Viel}}\ and\ \bibinfo {author} {\bibfnamefont {K.~B.}\ \bibnamefont
  {Whaley}},\ }\href@noop {} {\bibfield  {journal} {\bibinfo  {journal} {J.
  Chem. Phys.}\ }\textbf {\bibinfo {volume} {115}},\ \bibinfo {pages} {10186}
  (\bibinfo {year} {2001})}\BibitemShut {NoStop}%
\bibitem [{\citenamefont {McCoy}(2006)}]{mccoy2006}%
  \BibitemOpen
  \bibfield  {author} {\bibinfo {author} {\bibfnamefont {A.~B.}\ \bibnamefont
  {McCoy}},\ }\href@noop {} {\bibfield  {journal} {\bibinfo  {journal} {Inter.
  Rev. Phys. Chem.}\ }\textbf {\bibinfo {volume} {25}},\ \bibinfo {pages} {77}
  (\bibinfo {year} {2006})}\BibitemShut {NoStop}%
\bibitem [{\citenamefont {Austin}, \citenamefont {Zubarev},\ and\ \citenamefont
  {Lester~Jr}(2011)}]{austin2011}%
  \BibitemOpen
  \bibfield  {author} {\bibinfo {author} {\bibfnamefont {B.~M.}\ \bibnamefont
  {Austin}}, \bibinfo {author} {\bibfnamefont {D.~Y.}\ \bibnamefont {Zubarev}},
  \ and\ \bibinfo {author} {\bibfnamefont {W.~A.}\ \bibnamefont {Lester~Jr}},\
  }\href@noop {} {\bibfield  {journal} {\bibinfo  {journal} {Chem. Rev.}\
  }\textbf {\bibinfo {volume} {112}},\ \bibinfo {pages} {263} (\bibinfo {year}
  {2011})}\BibitemShut {NoStop}%
\bibitem [{\citenamefont {Petit}\ and\ \citenamefont
  {McCoy}(2013)}]{mccoy2013_2}%
  \BibitemOpen
  \bibfield  {author} {\bibinfo {author} {\bibfnamefont {A.~S.}\ \bibnamefont
  {Petit}}\ and\ \bibinfo {author} {\bibfnamefont {A.~B.}\ \bibnamefont
  {McCoy}},\ }\href@noop {} {\bibfield  {journal} {\bibinfo  {journal} {J.
  Phys. Chem. A}\ }\textbf {\bibinfo {volume} {117}},\ \bibinfo {pages} {7009}
  (\bibinfo {year} {2013})}\BibitemShut {NoStop}%
\bibitem [{\citenamefont {Jones}\ \emph {et~al.}(2009)\citenamefont {Jones},
  \citenamefont {Thompson}, \citenamefont {Crain}, \citenamefont {M{\"u}ser},\
  and\ \citenamefont {Martyna}}]{jones2009}%
  \BibitemOpen
  \bibfield  {author} {\bibinfo {author} {\bibfnamefont {A.}~\bibnamefont
  {Jones}}, \bibinfo {author} {\bibfnamefont {A.}~\bibnamefont {Thompson}},
  \bibinfo {author} {\bibfnamefont {J.}~\bibnamefont {Crain}}, \bibinfo
  {author} {\bibfnamefont {M.~H.}\ \bibnamefont {M{\"u}ser}}, \ and\ \bibinfo
  {author} {\bibfnamefont {G.~J.}\ \bibnamefont {Martyna}},\ }\href@noop {}
  {\bibfield  {journal} {\bibinfo  {journal} {Phys. Rev. B}\ }\textbf {\bibinfo
  {volume} {79}},\ \bibinfo {pages} {144119} (\bibinfo {year}
  {2009})}\BibitemShut {NoStop}%
\bibitem [{\citenamefont {Babin}\ and\ \citenamefont
  {Paesani}(2013)}]{babin2013}%
  \BibitemOpen
  \bibfield  {author} {\bibinfo {author} {\bibfnamefont {V.}~\bibnamefont
  {Babin}}\ and\ \bibinfo {author} {\bibfnamefont {F.}~\bibnamefont
  {Paesani}},\ }\href@noop {} {\bibfield  {journal} {\bibinfo  {journal} {Chem.
  Phys. Lett.}\ }\textbf {\bibinfo {volume} {580}},\ \bibinfo {pages} {1}
  (\bibinfo {year} {2013})}\BibitemShut {NoStop}%
\bibitem [{\citenamefont {Boninsegni}\ and\ \citenamefont
  {Moroni}(2012)}]{boninsegni2012}%
  \BibitemOpen
  \bibfield  {author} {\bibinfo {author} {\bibfnamefont {M.}~\bibnamefont
  {Boninsegni}}\ and\ \bibinfo {author} {\bibfnamefont {S.}~\bibnamefont
  {Moroni}},\ }\href@noop {} {\bibfield  {journal} {\bibinfo  {journal} {Phys.
  Rev. E}\ }\textbf {\bibinfo {volume} {86}},\ \bibinfo {pages} {056712}
  (\bibinfo {year} {2012})}\BibitemShut {NoStop}%
\bibitem [{\citenamefont {Cuervo}, \citenamefont {Roy},\ and\ \citenamefont
  {Boninsegni}(2005)}]{cuervo2005}%
  \BibitemOpen
  \bibfield  {author} {\bibinfo {author} {\bibfnamefont {J.~E.}\ \bibnamefont
  {Cuervo}}, \bibinfo {author} {\bibfnamefont {P.-N.}\ \bibnamefont {Roy}}, \
  and\ \bibinfo {author} {\bibfnamefont {M.}~\bibnamefont {Boninsegni}},\
  }\href@noop {} {\bibfield  {journal} {\bibinfo  {journal} {J. Chem. Phys.}\
  }\textbf {\bibinfo {volume} {122}},\ \bibinfo {pages} {114504} (\bibinfo
  {year} {2005})}\BibitemShut {NoStop}%
\bibitem [{\citenamefont {Guardiola}\ and\ \citenamefont
  {Navarro}(2008)}]{guardiola2008}%
  \BibitemOpen
  \bibfield  {author} {\bibinfo {author} {\bibfnamefont {R.}~\bibnamefont
  {Guardiola}}\ and\ \bibinfo {author} {\bibfnamefont {J.}~\bibnamefont
  {Navarro}},\ }\href@noop {} {\bibfield  {journal} {\bibinfo  {journal} {Cent.
  Euro. J. Phys.}\ }\textbf {\bibinfo {volume} {6}},\ \bibinfo {pages} {33}
  (\bibinfo {year} {2008})}\BibitemShut {NoStop}%
\bibitem [{\citenamefont {Sola}\ and\ \citenamefont
  {Boronat}(2011)}]{sola2011}%
  \BibitemOpen
  \bibfield  {author} {\bibinfo {author} {\bibfnamefont {E.}~\bibnamefont
  {Sola}}\ and\ \bibinfo {author} {\bibfnamefont {J.}~\bibnamefont {Boronat}},\
  }\href@noop {} {\bibfield  {journal} {\bibinfo  {journal} {J. Phys. Chem. A}\
  }\textbf {\bibinfo {volume} {115}},\ \bibinfo {pages} {7071} (\bibinfo {year}
  {2011})}\BibitemShut {NoStop}%
\bibitem [{\citenamefont {Nemec}(2010)}]{nemec2010}%
  \BibitemOpen
  \bibfield  {author} {\bibinfo {author} {\bibfnamefont {N.}~\bibnamefont
  {Nemec}},\ }\href@noop {} {\bibfield  {journal} {\bibinfo  {journal} {Phys.
  Rev. B}\ }\textbf {\bibinfo {volume} {81}},\ \bibinfo {pages} {035119}
  (\bibinfo {year} {2010})}\BibitemShut {NoStop}%
\bibitem [{\citenamefont {Jiang}\ \emph {et~al.}(2005)\citenamefont {Jiang},
  \citenamefont {Xu}, \citenamefont {Hutson},\ and\ \citenamefont
  {Ba{\v{c}}i{\'c}}}]{bacic2005}%
  \BibitemOpen
  \bibfield  {author} {\bibinfo {author} {\bibfnamefont {H.}~\bibnamefont
  {Jiang}}, \bibinfo {author} {\bibfnamefont {M.}~\bibnamefont {Xu}}, \bibinfo
  {author} {\bibfnamefont {J.~M.}\ \bibnamefont {Hutson}}, \ and\ \bibinfo
  {author} {\bibfnamefont {Z.}~\bibnamefont {Ba{\v{c}}i{\'c}}},\ }\href@noop {}
  {\bibfield  {journal} {\bibinfo  {journal} {J. Chem. Phys.}\ }\textbf
  {\bibinfo {volume} {123}},\ \bibinfo {pages} {054305} (\bibinfo {year}
  {2005})}\BibitemShut {NoStop}%
\bibitem [{\citenamefont {Wang}\ \emph {et~al.}(2012)\citenamefont {Wang},
  \citenamefont {Babin}, \citenamefont {Bowman},\ and\ \citenamefont
  {Paesani}}]{wang2012}%
  \BibitemOpen
  \bibfield  {author} {\bibinfo {author} {\bibfnamefont {Y.}~\bibnamefont
  {Wang}}, \bibinfo {author} {\bibfnamefont {V.}~\bibnamefont {Babin}},
  \bibinfo {author} {\bibfnamefont {J.~M.}\ \bibnamefont {Bowman}}, \ and\
  \bibinfo {author} {\bibfnamefont {F.}~\bibnamefont {Paesani}},\ }\href@noop
  {} {\bibfield  {journal} {\bibinfo  {journal} {J. Am. Chem. Soc.}\ }\textbf
  {\bibinfo {volume} {134}},\ \bibinfo {pages} {11116} (\bibinfo {year}
  {2012})}\BibitemShut {NoStop}%
\bibitem [{\citenamefont {Severson}\ and\ \citenamefont
  {Buch}(1999)}]{buch1999}%
  \BibitemOpen
  \bibfield  {author} {\bibinfo {author} {\bibfnamefont {M.~W.}\ \bibnamefont
  {Severson}}\ and\ \bibinfo {author} {\bibfnamefont {V.}~\bibnamefont
  {Buch}},\ }\href@noop {} {\bibfield  {journal} {\bibinfo  {journal} {J. Chem.
  Phys.}\ }\textbf {\bibinfo {volume} {111}},\ \bibinfo {pages} {10866}
  (\bibinfo {year} {1999})}\BibitemShut {NoStop}%
\bibitem [{\citenamefont {Severson}, \citenamefont {Devlin},\ and\
  \citenamefont {Buch}(2003)}]{buch2003}%
  \BibitemOpen
  \bibfield  {author} {\bibinfo {author} {\bibfnamefont {M.~W.}\ \bibnamefont
  {Severson}}, \bibinfo {author} {\bibfnamefont {J.~P.}\ \bibnamefont
  {Devlin}}, \ and\ \bibinfo {author} {\bibfnamefont {V.}~\bibnamefont
  {Buch}},\ }\href@noop {} {\bibfield  {journal} {\bibinfo  {journal} {J. Chem.
  Phys.}\ }\textbf {\bibinfo {volume} {119}},\ \bibinfo {pages} {4449}
  (\bibinfo {year} {2003})}\BibitemShut {NoStop}%
\bibitem [{\citenamefont {Goldman}\ and\ \citenamefont
  {Saykally}(2004)}]{goldman2004}%
  \BibitemOpen
  \bibfield  {author} {\bibinfo {author} {\bibfnamefont {N.}~\bibnamefont
  {Goldman}}\ and\ \bibinfo {author} {\bibfnamefont {R.}~\bibnamefont
  {Saykally}},\ }\href@noop {} {\bibfield  {journal} {\bibinfo  {journal} {J.
  Chem. Phys.}\ }\textbf {\bibinfo {volume} {120}},\ \bibinfo {pages} {4777}
  (\bibinfo {year} {2004})}\BibitemShut {NoStop}%
\bibitem [{\citenamefont {Gillan}\ \emph {et~al.}(2013)\citenamefont {Gillan},
  \citenamefont {Alf{\`e}}, \citenamefont {Bart{\'o}k},\ and\ \citenamefont
  {Cs{\'a}nyi}}]{gillan2013}%
  \BibitemOpen
  \bibfield  {author} {\bibinfo {author} {\bibfnamefont {M.}~\bibnamefont
  {Gillan}}, \bibinfo {author} {\bibfnamefont {D.}~\bibnamefont {Alf{\`e}}},
  \bibinfo {author} {\bibfnamefont {A.}~\bibnamefont {Bart{\'o}k}}, \ and\
  \bibinfo {author} {\bibfnamefont {G.}~\bibnamefont {Cs{\'a}nyi}},\
  }\href@noop {} {\bibfield  {journal} {\bibinfo  {journal} {J. Chem. Phys.}\
  }\textbf {\bibinfo {volume} {139}},\ \bibinfo {pages} {244504} (\bibinfo
  {year} {2013})}\BibitemShut {NoStop}%
\bibitem [{\citenamefont {Gregory}\ and\ \citenamefont
  {Clary}(1996)}]{gregory1996_1}%
  \BibitemOpen
  \bibfield  {author} {\bibinfo {author} {\bibfnamefont {J.~K.}\ \bibnamefont
  {Gregory}}\ and\ \bibinfo {author} {\bibfnamefont {D.~C.}\ \bibnamefont
  {Clary}},\ }\href@noop {} {\bibfield  {journal} {\bibinfo  {journal} {J.
  Phys. Chem.}\ }\textbf {\bibinfo {volume} {100}},\ \bibinfo {pages} {18014}
  (\bibinfo {year} {1996})}\BibitemShut {NoStop}%
\bibitem [{\citenamefont {Sorenson}, \citenamefont {Gregory},\ and\
  \citenamefont {Clary}(1996)}]{gregory1996_2}%
  \BibitemOpen
  \bibfield  {author} {\bibinfo {author} {\bibfnamefont {J.~M.}\ \bibnamefont
  {Sorenson}}, \bibinfo {author} {\bibfnamefont {J.~K.}\ \bibnamefont
  {Gregory}}, \ and\ \bibinfo {author} {\bibfnamefont {D.~C.}\ \bibnamefont
  {Clary}},\ }\href@noop {} {\bibfield  {journal} {\bibinfo  {journal} {Chem.
  Phys. Lett.}\ }\textbf {\bibinfo {volume} {263}},\ \bibinfo {pages} {680}
  (\bibinfo {year} {1996})}\BibitemShut {NoStop}%
\bibitem [{\citenamefont {Liu}\ \emph {et~al.}(1996)\citenamefont {Liu},
  \citenamefont {Brown}, \citenamefont {Carter}, \citenamefont {Saykally},
  \citenamefont {Gregory},\ and\ \citenamefont {Clary}}]{clary1996}%
  \BibitemOpen
  \bibfield  {author} {\bibinfo {author} {\bibfnamefont {K.}~\bibnamefont
  {Liu}}, \bibinfo {author} {\bibfnamefont {M.}~\bibnamefont {Brown}}, \bibinfo
  {author} {\bibfnamefont {C.}~\bibnamefont {Carter}}, \bibinfo {author}
  {\bibfnamefont {R.}~\bibnamefont {Saykally}}, \bibinfo {author}
  {\bibfnamefont {J.}~\bibnamefont {Gregory}}, \ and\ \bibinfo {author}
  {\bibfnamefont {D.}~\bibnamefont {Clary}},\ }\href@noop {} {\bibfield
  {journal} {\bibinfo  {journal} {Nature}\ }\textbf {\bibinfo {volume} {381}},\
  \bibinfo {pages} {501} (\bibinfo {year} {1996})}\BibitemShut {NoStop}%
\bibitem [{\citenamefont {Assaraf}, \citenamefont {Caffarel},\ and\
  \citenamefont {Khelif}(2000)}]{assaraf2000}%
  \BibitemOpen
  \bibfield  {author} {\bibinfo {author} {\bibfnamefont {R.}~\bibnamefont
  {Assaraf}}, \bibinfo {author} {\bibfnamefont {M.}~\bibnamefont {Caffarel}}, \
  and\ \bibinfo {author} {\bibfnamefont {A.}~\bibnamefont {Khelif}},\
  }\href@noop {} {\bibfield  {journal} {\bibinfo  {journal} {Phys. Rev. E}\
  }\textbf {\bibinfo {volume} {61}},\ \bibinfo {pages} {4566} (\bibinfo {year}
  {2000})}\BibitemShut {NoStop}%
\bibitem [{\citenamefont {Suhm}\ and\ \citenamefont {Watts}(1991)}]{watts1991}%
  \BibitemOpen
  \bibfield  {author} {\bibinfo {author} {\bibfnamefont {M.~A.}\ \bibnamefont
  {Suhm}}\ and\ \bibinfo {author} {\bibfnamefont {R.~O.}\ \bibnamefont
  {Watts}},\ }\href@noop {} {\bibfield  {journal} {\bibinfo  {journal} {Phys.
  Rep.}\ }\textbf {\bibinfo {volume} {204}},\ \bibinfo {pages} {293} (\bibinfo
  {year} {1991})}\BibitemShut {NoStop}%
\bibitem [{\citenamefont {Umrigar}, \citenamefont {Nightingale},\ and\
  \citenamefont {Runge}(1993)}]{umrigar1993}%
  \BibitemOpen
  \bibfield  {author} {\bibinfo {author} {\bibfnamefont {C.}~\bibnamefont
  {Umrigar}}, \bibinfo {author} {\bibfnamefont {M.}~\bibnamefont
  {Nightingale}}, \ and\ \bibinfo {author} {\bibfnamefont {K.}~\bibnamefont
  {Runge}},\ }\href@noop {} {\bibfield  {journal} {\bibinfo  {journal} {J.
  Chem. Phys.}\ }\textbf {\bibinfo {volume} {99}},\ \bibinfo {pages} {2865}
  (\bibinfo {year} {1993})}\BibitemShut {NoStop}%
\bibitem [{\citenamefont {Warren}\ and\ \citenamefont
  {Hinde}(2006)}]{warren2006}%
  \BibitemOpen
  \bibfield  {author} {\bibinfo {author} {\bibfnamefont {G.~L.}\ \bibnamefont
  {Warren}}\ and\ \bibinfo {author} {\bibfnamefont {R.~J.}\ \bibnamefont
  {Hinde}},\ }\href@noop {} {\bibfield  {journal} {\bibinfo  {journal} {Phys.
  Rev. E}\ }\textbf {\bibinfo {volume} {73}},\ \bibinfo {pages} {056706}
  (\bibinfo {year} {2006})}\BibitemShut {NoStop}%
\bibitem [{\citenamefont {Hetherington}(1984)}]{hetherington1984}%
  \BibitemOpen
  \bibfield  {author} {\bibinfo {author} {\bibfnamefont {J.}~\bibnamefont
  {Hetherington}},\ }\href@noop {} {\bibfield  {journal} {\bibinfo  {journal}
  {Phys. Rev. A}\ }\textbf {\bibinfo {volume} {30}},\ \bibinfo {pages} {2713}
  (\bibinfo {year} {1984})}\BibitemShut {NoStop}%
\bibitem [{\citenamefont {Habershon}, \citenamefont {Markland},\ and\
  \citenamefont {Manolopoulos}(2009)}]{habershon2009}%
  \BibitemOpen
  \bibfield  {author} {\bibinfo {author} {\bibfnamefont {S.}~\bibnamefont
  {Habershon}}, \bibinfo {author} {\bibfnamefont {T.~E.}\ \bibnamefont
  {Markland}}, \ and\ \bibinfo {author} {\bibfnamefont {D.~E.}\ \bibnamefont
  {Manolopoulos}},\ }\href@noop {} {\bibfield  {journal} {\bibinfo  {journal}
  {J. Chem. Phys.}\ }\textbf {\bibinfo {volume} {131}},\ \bibinfo {pages}
  {024501} (\bibinfo {year} {2009})}\BibitemShut {NoStop}%
\bibitem [{\citenamefont {Wang}\ \emph {et~al.}(2011)\citenamefont {Wang},
  \citenamefont {Huang}, \citenamefont {Shepler}, \citenamefont {Braams},\ and\
  \citenamefont {Bowman}}]{wang2011}%
  \BibitemOpen
  \bibfield  {author} {\bibinfo {author} {\bibfnamefont {Y.}~\bibnamefont
  {Wang}}, \bibinfo {author} {\bibfnamefont {X.}~\bibnamefont {Huang}},
  \bibinfo {author} {\bibfnamefont {B.~C.}\ \bibnamefont {Shepler}}, \bibinfo
  {author} {\bibfnamefont {B.~J.}\ \bibnamefont {Braams}}, \ and\ \bibinfo
  {author} {\bibfnamefont {J.~M.}\ \bibnamefont {Bowman}},\ }\href@noop {}
  {\bibfield  {journal} {\bibinfo  {journal} {J. Chem. Phys.}\ }\textbf
  {\bibinfo {volume} {134}},\ \bibinfo {pages} {094509} (\bibinfo {year}
  {2011})}\BibitemShut {NoStop}%
\bibitem [{\citenamefont {Medders}, \citenamefont {Babin},\ and\ \citenamefont
  {Paesani}(2013)}]{medders2013}%
  \BibitemOpen
  \bibfield  {author} {\bibinfo {author} {\bibfnamefont {G.~R.}\ \bibnamefont
  {Medders}}, \bibinfo {author} {\bibfnamefont {V.}~\bibnamefont {Babin}}, \
  and\ \bibinfo {author} {\bibfnamefont {F.}~\bibnamefont {Paesani}},\
  }\href@noop {} {\bibfield  {journal} {\bibinfo  {journal} {J. Chem. Theory
  and Comp.}\ }\textbf {\bibinfo {volume} {9}},\ \bibinfo {pages} {1103}
  (\bibinfo {year} {2013})}\BibitemShut {NoStop}%
\bibitem [{\citenamefont {Babin}, \citenamefont {Medders},\ and\ \citenamefont
  {Paesani}(2014)}]{babin2014}%
  \BibitemOpen
  \bibfield  {author} {\bibinfo {author} {\bibfnamefont {V.}~\bibnamefont
  {Babin}}, \bibinfo {author} {\bibfnamefont {G.~R.}\ \bibnamefont {Medders}},
  \ and\ \bibinfo {author} {\bibfnamefont {F.}~\bibnamefont {Paesani}},\
  }\href@noop {} {\bibfield  {journal} {\bibinfo  {journal} {J. Chem. Theory
  and Comp.}\ }\textbf {\bibinfo {volume} {10}},\ \bibinfo {pages} {1599}
  (\bibinfo {year} {2014})}\BibitemShut {NoStop}%
\bibitem [{\citenamefont {Lin}\ and\ \citenamefont
  {McCoy}(2013)}]{mccoy2013_1}%
  \BibitemOpen
  \bibfield  {author} {\bibinfo {author} {\bibfnamefont {Z.}~\bibnamefont
  {Lin}}\ and\ \bibinfo {author} {\bibfnamefont {A.~B.}\ \bibnamefont
  {McCoy}},\ }\href@noop {} {\bibfield  {journal} {\bibinfo  {journal} {J.
  Phys. Chem. A}\ }\textbf {\bibinfo {volume} {117}},\ \bibinfo {pages} {11725}
  (\bibinfo {year} {2013})}\BibitemShut {NoStop}%
\bibitem [{\citenamefont {Acioli}\ \emph {et~al.}(2008)\citenamefont {Acioli},
  \citenamefont {Xie}, \citenamefont {Braams},\ and\ \citenamefont
  {Bowman}}]{bowman2008}%
  \BibitemOpen
  \bibfield  {author} {\bibinfo {author} {\bibfnamefont {P.~H.}\ \bibnamefont
  {Acioli}}, \bibinfo {author} {\bibfnamefont {Z.}~\bibnamefont {Xie}},
  \bibinfo {author} {\bibfnamefont {B.~J.}\ \bibnamefont {Braams}}, \ and\
  \bibinfo {author} {\bibfnamefont {J.~M.}\ \bibnamefont {Bowman}},\
  }\href@noop {} {\bibfield  {journal} {\bibinfo  {journal} {J. Chem. Phys.}\
  }\textbf {\bibinfo {volume} {128}},\ \bibinfo {pages} {104318} (\bibinfo
  {year} {2008})}\BibitemShut {NoStop}%
\bibitem [{\citenamefont {Rocher-Casterline}\ \emph {et~al.}(2011)\citenamefont
  {Rocher-Casterline}, \citenamefont {Ch'ng}, \citenamefont {Mollner},\ and\
  \citenamefont {Reisler}}]{reisler2011}%
  \BibitemOpen
  \bibfield  {author} {\bibinfo {author} {\bibfnamefont {B.~E.}\ \bibnamefont
  {Rocher-Casterline}}, \bibinfo {author} {\bibfnamefont {L.~C.}\ \bibnamefont
  {Ch'ng}}, \bibinfo {author} {\bibfnamefont {A.~K.}\ \bibnamefont {Mollner}},
  \ and\ \bibinfo {author} {\bibfnamefont {H.}~\bibnamefont {Reisler}},\
  }\href@noop {} {\bibfield  {journal} {\bibinfo  {journal} {J. Chem. Phys.}\
  }\textbf {\bibinfo {volume} {134}},\ \bibinfo {pages} {211101} (\bibinfo
  {year} {2011})}\BibitemShut {NoStop}%
\bibitem [{\citenamefont {Ch’ng}\ \emph {et~al.}(2012)\citenamefont
  {Ch’ng}, \citenamefont {Samanta}, \citenamefont {Czakó}, \citenamefont
  {Bowman},\ and\ \citenamefont {Reisler}}]{reisler2012}%
  \BibitemOpen
  \bibfield  {author} {\bibinfo {author} {\bibfnamefont {L.~C.}\ \bibnamefont
  {Ch’ng}}, \bibinfo {author} {\bibfnamefont {A.~K.}\ \bibnamefont
  {Samanta}}, \bibinfo {author} {\bibfnamefont {G.}~\bibnamefont {Czakó}},
  \bibinfo {author} {\bibfnamefont {J.~M.}\ \bibnamefont {Bowman}}, \ and\
  \bibinfo {author} {\bibfnamefont {H.}~\bibnamefont {Reisler}},\ }\href@noop
  {} {\bibfield  {journal} {\bibinfo  {journal} {J. Am. Chem. Soc.}\ }\textbf
  {\bibinfo {volume} {134}},\ \bibinfo {pages} {15430} (\bibinfo {year}
  {2012})}\BibitemShut {NoStop}%
\bibitem [{\citenamefont {Brown}\ and\ \citenamefont
  {Mandelshtam}(2014)}]{brown2014}%
  \BibitemOpen
  \bibfield  {author} {\bibinfo {author} {\bibfnamefont {S.~E.}\ \bibnamefont
  {Brown}}\ and\ \bibinfo {author} {\bibfnamefont {V.~A.}\ \bibnamefont
  {Mandelshtam}},\ }\href@noop {} {\bibfield  {journal} {\bibinfo  {journal}
  {In Progress}\ } (\bibinfo {year} {2014})}\BibitemShut {NoStop}%
\end{thebibliography}
%
\end{document}